\documentclass[prd,showkeys,floatfix,nofootinbib,
               preprint,12pt,
               tightenlines,fleqn]{revtex4} 

\usepackage{amsmath,amssymb,revsymb,graphicx,dcolumn}
\usepackage{leftidx}

\usepackage{slashed}
\usepackage[usenames, dvipsnames]{color}
 
\definecolor{new}{rgb}{0.858, 0.188, 0.478}

\newcommand{\cev}[1]{\reflectbox{\ensuremath{\vec{\reflectbox{\ensuremath{#1}}}}}}

\newcommand{\beq}{\begin{equation}}
\newcommand{\eeq}{\end{equation}}
\newcommand{\beqa}{\begin{eqnarray}}
\newcommand{\eeqa}{\end{eqnarray}}
\newcommand{\bsubeqs}{\begin{subequations}}
\newcommand{\esubeqs}{\end{subequations}}

\usepackage{mwe}
\newcommand{\imineq}[2]{\vcenter{\hbox{\includegraphics[height=#2ex]{#1}}}}

\begin{document}
\hfill KA--TP--25--2016 \vspace*{8mm}\newline
\title[]
      {Near-horizon physics of an evaporating black hole: \\ One-loop effects in the $\lambda\Phi^4$-theory}
\author{Slava Emelyanov}
\email{viacheslav.emelyanov@kit.edu}
\affiliation{Institute for Theoretical Physics,\\
Karlsruhe Institute of Technology (KIT),\\
76131 Karlsruhe, Germany\\}

\begin{abstract}
\vspace*{2.5mm}\noindent
We study massless scalar theory with the quartic self-interacting term far away from and near to
evaporating and spherically symmetric black hole. We propose a principle of how to define the
physical notion of particle in curved space-time. Employing this definition, we compute one-loop correction
to the self-energy and coupling constant of the scalar field near the horizon in the freely-falling frame. 
We find that the coupling constant becomes slightly stronger near to the horizon. We
also find that the term
in the 2-point function that is (partially) responsible for the black-hole evaporation corresponds to the
sub-leading correction to the Feynman propagator whose pole structure is of the leading order.
\end{abstract}


\maketitle

\section{Introduction}

The quantum physics of black holes is full of various paradoxes.
One of these problems is related to the understanding of the particle production
process induced by black holes. The created particles $\{|\tilde{\psi}_i\rangle\}$ possess a thermal profile and 
turn out to appear in empty space, i.e. outside of the matter that has gravitationally collapsed into a black hole.
The fact is that such a process 
is in apparent conflict with the unitary evolution in quantum theory~\cite{Hawking}. In other words, 
the $S$-matrix between initial vacuum state and final thermal state cannot be unitary. If one 
demands the $S$-matrix be unitary, one comes to another problem - the firewall
paradox~\cite{Almheiri&Marolf&Polchinski&Sully}, - i.e. the near-horizon region is populated by the 
high-energy particles of the type $\{|\tilde{\psi}_i\rangle\}$. If this was correct, then this would imply
the equivalence principle, being at the heart of general relativity, would not in general hold in quantum
theory. 

There exists a logically non-excludable interpretation of the black-hole evaporation (non-vanishing 
positive outgoing energy flux) which seems to be free of the mentioned inconsistencies.
The quantum state remains empty outside of the collapsed matter, i.e. there are no physical particles in
the locally Minkowski
vacuum (we denote this vacuum as $|\Omega\rangle$ in the following). In other words, the initial 
Hilbert space representation $\mathcal{H}$ of the algebra $\mathcal{A}$ of all field operators known 
in the Standard Model is still a physical representation even after the black hole has formed. 
This implies in particular that the spectrum of particles $\{|\psi_i\rangle\}$ as being elements of 
$\mathcal{H}$ (assigned far from the event horizon to the irreducible representations of the 
Poincar\'{e} group $\mathcal{P}_+^\uparrow$; below we show how this can be done near the 
event horizon) from which the collapsing matter were composed are still physical excitations after 
the gravitational collapse.

The crucial question is then whether thermally distributed particles $\{|\tilde{\psi}_i\rangle\}$
are elements of $\mathcal{H}$. To put it differently, the question is whether there exist projection operators 
$\hat{P}_i \in \mathcal{A}$ which map $\mathcal{H}$ on a one-dimensional subspace associated with
each of $|\tilde{\psi}_i\rangle$. The fact, however, is that these projections do not exist in 
$\mathcal{A}$ and, hence, the particles $\{|\tilde{\psi}_i\rangle\}$ are elements of another Hilbert 
space $\tilde{\mathcal{H}} \not\subset \mathcal{H}$~\cite{Emelyanov-2015-2}. This can be shown by 
employing the argument that the splitting of the algebra $\mathcal{A}$ into factor subalgebras does not lead 
to the factorization of the Hilbert space $\mathcal{H}$ in quantum field theory (in contrast to quantum 
mechanics with finite dimensional Hilbert space representations, e.g., of a finite qubit system, which are all 
unitarily equivalent according to the Stone-von Neumann theorem). Thus, there does not exist a 
unitary $S$-matrix mapping elements of $\mathcal{H}$ into elements of $\tilde{\mathcal{H}}$ within the 
framework of local quantum field theory. Therefore, for the states $\{|\tilde{\psi}_i\rangle\}$
to be physically realisable as
quanta of the Standard Model fields, there must occur a phase transition
$\mathcal{H} \rightarrow \tilde{\mathcal{H}}$ 
whenever a black hole forms. This scenario is physically unacceptable.

How can the black-hole evaporation then be understood? The evaporation process can still be accounted 
for within local quantum field theory in a consistent way without referring to the states
$\{|\tilde{\psi}_i\rangle\}$. The 
non-trivial value of the renormalised stress tensor $\langle \hat{T}_\nu^\mu\rangle$ of a certain quantum 
field is assigned to the vacuum $|\Omega\rangle$. In other words, the vacuum is gravitationally ``active" 
after the event horizon has formed (as $\langle \hat{T}_\nu^\mu\rangle$ enters the Einstein equations). 
This occurs, because the field operators are sensitive (through their field equations) to
the geometry of
space-time and, hence, are modified not too far away from the horizon. The modification of the field
operators results in the change of the vacuum stress tensor $\langle \hat{T}_\nu^\mu\rangle$ (this 
explains the featurelessness of the outgoing energy flux as the state $|\Omega\rangle$ is full of the 
featureless quantum fluctuations only). This tensor decreases as $1/r^2$ far from the 
hole~\cite{Christensen&Fulling,Candelas} and is thus practically zero sufficiently far away from
the horizon. 

An analogous effect occurs, e.g., in the Casimir set-up, where the (Minkowski) vacuum $|\Omega\rangle$ 
also possesses a non-vanishing energy-momentum tensor. This is due to the modification of the 
electromagnetic operators $\hat{E}_i$ and $\hat{B}_i$ between the conducting plates. This leads to 
$\langle\hat{E}_i\hat{E}_j\rangle \neq 0$ and $\langle\hat{B}_i\hat{B}_j\rangle \neq 0$, although these 
vanish in the absence of the plates.

This analogy seems also to hold when one studies electromagnetic properties of the vacuum within quantum 
electrodynamics. The electromagnetic properties of the vacuum $|\Omega\rangle$ are characterised by 
the electric permittivity $\epsilon$ and the magnetic permeability $\mu$ entering the Maxwell equations. 
It has been found within the Casimir set-up that the dispersion relation of low-energy
photons is modified in-between the plates~\cite{Scharnhorst}. We have recently shown that a similar effect
exists in the black-hole background~\cite{Emelyanov-2016-1}. Moreover, we have shown that photons
acquire an effective mass decreasing as $1/r$ and a point-like electric charge can be partially screened due
to black holes~\cite{Emelyanov-2016-2}.

In this paper we study properties of the vacuum $|\Omega\rangle$ in the framework of the
massless $\lambda\Phi^4$-theory. The main purpose is to find out any physical inconsistencies related to
the idea of having medium-like characteristics of the vacuum in the background of black holes. Specifically,
we consider a self-interacting scalar model in the far-from- and near-horizon
region of a large black hole. The
one-loop correction to the self-energy and coupling constant are computed in the near-horizon
region. As expected, we find that these corrections are suppressed as $(\lambda_\mathbf{p}/r_H)^2$ near the
event horizon, where $\lambda_\mathbf{p}$ is the de Broglie wavelength of the scalar particle and $r_H$ is
a size of the black-hole horizon.

The outline of this paper is as follows. In Sec.~\ref{sec:physical-particles}, we introduce the principle of
how the physical particles should be defined in curved space-times. The basic idea is to employ a local 
Minkowski frame to identify a particle with a localised state as one has been successfully
doing that in 
Minkowski space. The equivalence principle plays a crucial role in extending the particle notion to any 
space-time domain in which gravity is not too strong. This tacitly implies that the notion of particle
is \emph{not} observer-dependent, i.e. covariant, as opposed to the common belief.

In Sec.~\ref{sec:scalar-model},
the self-interacting scalar field is considered in the background of an evaporating and
spherically symmetric black hole of astrophysical mass, i.e. $M \geq M_\odot$, where $M_\odot$ is the
solar mass. For these black holes, the local inertial frame might be of the size
$l_M \gtrsim 300(M/M_\odot)\,\text{m}$. Thus, a particle detector of the size $l_D \ll 30(M/M_\odot)\,\text{m}$
could be employed to study how particle scattering reactions quantitatively differ from the
same reactions
in the asymptotically flat region, i.e. in the region, where the influence of black holes can be ignored.

In Sec.~\ref{sec:lpp}, we provide an argument why the firewall paradox does not exist.
The non-existence of the unitary $S$-matrix relating the states $\{|\psi_i\rangle\}$ with
$\{|\tilde{\psi}_i\rangle\}$ is on the contrary consistent with local quantum field theory, but does not imply
the unitarity in the gravitational collapse is broken. The outstanding problem is how to take into account
the backreaction of the quantum fields on the quantum state of the collapsed matter which is under the 
horizon.

In Sec.~\ref{sec:concluding remarks}, the main results are summarised.

Throughout this paper the fundamental constants are set to $c = G = k_\text{B} = \hbar = 1$, unless
stated otherwise.

\section{Local notion of particle in curved space-time}
\label{sec:physical-particles}

We want to study certain scattering processes in the vicinity of the horizon of a large black hole
formed through the gravitational collapse and compare these with observations in the far-from-horizon
region. For this purpose it is first necessary to define a physical notion of particle in curved
space-time. Bearing in mind the remarkable success of particle physics based on QFT, 
the notion of particle should be related to the pole structure of the Feynman propagator when computed
in the local Minkowski frame. Therefore, the guiding principle should be based on reproducing the standard
results of particle physics at any given point in the universe.

Particle physics formulated in Minkowski space and based on the Standard Model 
has successfully passed so far all tests in the particle colliders up to the energy scale $1\,\text{TeV}$.
Certainly, there is physics beyond the Standard Model which is assumed to be associated with
the theory itself (e.g., the neutrino oscillation implies that
at least two among of three neutrino flavors are massive), rather than the modifications of
the basic QFT principles.

However, the universe is globally non-flat. The observable part of the universe looks
at cosmological scales
($100\,\text{Mpc} \lesssim l \lesssim 3000\,\text{Mpc}$) as de Sitter space due to dark energy (see,
e.g.,~\cite{Mukhanov}). The universe becomes inhomogeneous and anisotropic at smaller scales due
to dark matter, clusters of galaxies, galaxies and so on. At much smaller, but still macroscopic scales,
the universe definitely looks as being nearly Minkowski space.

Moreover, earth is also a source of the non-trivial local curvature.
If one introduces the normal Riemannian coordinates $y$, then the local geometry becomes flat:
\beqa
g_{\mu\nu}(y) &=& \eta_{\mu\nu}  + \frac{1}{3}R_{\mu\lambda\rho\nu}(y)\,y^{\lambda}y^{\rho} +
\text{O}\big(\nabla{R}\,y^3\big) \quad \text{with} \quad |Ry^2| \;\ll\; 1\,.
\eeqa
One usually employs the Minkowski-space approximation in order to describe various
scattering processes in the particle colliders. Therefore, it is \emph{a posteriori}
legitimate to use the Fourier-transform
technique (with integration over the whole space-time) whenever one restricts oneself to space-time
regions with a size $l \ll r(r/r_\oplus)^{\frac{1}{2}}$  ($r_\oplus \approx 8.7\,\text{mm}$ is earth's
gravitational radius) for the reason explained below. Specifically, the size $l_\text{LHC}$ of the
LHC is about $27\,\text{km}$ and we find 
$R_\oplus(R_\oplus/r_\oplus)^{\frac{1}{2}} \approx 1.7{\times}10^{8}\,\text{km} \gg R_\oplus \gg l_\text{LHC}$, 
where $R_\oplus \approx 6.4{\times}10^{3}\,\text{km}$ is earth's raduis. In other words, the gravitational
influence of earth on scattering processes in the LHC can be safely ignored. 

The integration over the whole space-time (as if it is infinitely large Minkowski space) in vertices in 
non-linear field theories does not entail any sort of inconsistencies,
because particles are described by localized states in
quantum field theory~\cite{Haag}. The particle state looks like the vacuum $|\Omega\rangle$
for measurements performed outside of its support. This is in turn characterised by the size of particle's wave 
packet. For instance, electron has a size which is about its Compton wavelength 
$\lambda_e = \frac{h}{m_ec} \approx 2.4{\times}10^{-12}\,\text{m}$. Thus, it makes a physical sense 
to speak about the electron (at rest) as a localised object, whenever one restricts oneself to space-time regions 
of the size $\lambda_e \ll l \ll l_c$, where $l_c$ is a characteristic size of the curvature 
(e.g., $l_c \sim 10^{8}\,\text{km}$ for earth and a hypothetical particle of the rest mass less than 
$10^{-17}\,\text{eV}$ could not be understood as being localised).

In Minkowski space-time, the wave packet $h_e(x)$ characterising the electron is a positive energy 
solution of the Dirac field equation. It possesses a non-vanishing support in the spatial
region of the extent $\lambda_e$. The notion of energy is defined with respect to the
Minkowski time 
translation operator $G  = \partial_\tau$ whose integral curves are geodesics, i.e. it satisfies the 
geodesic equation $\nabla_GG = 0$. Thus, one has
\beqa\label{eq:wave-packet}
h_e(x) &=& \frac{1}{\left(2\pi\right)^{3/2}}\int d^4p\;
\theta\big(p_0\big)\delta\big(p^2 - m_e^2\big)e^{-ipx}h_e(p)\,,
\eeqa
where $h_e(p)$ is a wave packet in momentum space and $x^0 = \tau$. At scales $l \ll l_c$ in curved 
space-time, the vector $G$ approximately satisfies the Killing equation, i.e. 
$l_c\nabla_{(\mu}G_{\nu)} \approx 0$, so that it is one of the generators of the \emph{local} Poincar\'{e} 
group. It turns out to be $l_c\nabla_{(\mu}G_{\nu)} \sim \text{O}(1)$ at larger scales $l \gtrsim l_c$, i.e.
$G$ is a local Killing vector.

Since the positive-energy packet $h(x)$ is chosen with respect to $G$ in particle physics and this choice
is consistent with the observations performed so far on earth (freely-moving, rather than moving along
any global Killing vector), we come to the following principle: \\[2mm]
\emph{A physical particle corresponds to a covariant wave packet $h(x)$ being a 
positive energy solution of the field equation with respect to a geodesic vector $G$ determining the 
dynamics in the local Minkowski frame.}\vspace{1.5mm}

The quantisation procedure of, e.g., a scalar non-interacting field $\hat{\Phi}(x)$ is performed by expanding 
the field over the positive- and negative-frequency modes defined with respect to a certain
time-like Killing vector $K$:
\beqa\label{eq:field-splitting}
\hat{\Phi}(x) &=& \int{}d\mu(\mathbf{p})\Big(\phi_\mathbf{p}(x)\,\hat{a}_\mathbf{p} +
\phi_\mathbf{p}^*(x)\,\hat{a}_\mathbf{p}^\dagger\Big) \;=\; \hat{a}(x) + \hat{a}^\dagger(x)\,,
\eeqa
where $K\phi_\mathbf{p} = - i\omega\phi_\mathbf{p}$ and 
$K\phi_\mathbf{p}^\dagger = + i\omega\phi_\mathbf{p}^\dagger$ and $d\mu(\mathbf{p})$ is some positive 
measure of integration/summation. Due to the linearity of the field equation, there are infinitely many ways
of choosing the modes $\phi_\mathbf{p}(x)$. According to our principle, there is a preferred
choice (unique up the local Lorentz transformation). This depends on a
point in curved space-time, such that $K = G$ and, hence, only locally satisfies the Killing equation.
In this case, $\hat{a}^\dagger(h)$ is a creation operator of the physical particle from the vacuum 
$|\Omega\rangle$ around of a support $\sigma$ of the wave packet $h(x)$. However, one has
\beqa\label{eq:particle-creation-operator}
\hat{a}^\dagger(h) &=& i\int_\Sigma d\Sigma_\mu\sqrt{-g(x)}\,g^{\mu\nu}(x)
\Big(\hat{\Phi}^\dagger(x)\nabla_\nu h(x) - h(x)\nabla_\nu\hat{\Phi}^\dagger(x)\Big)
\eeqa
for an arbitrary choice of the mode functions, where $\Sigma$ is a Cauchy surface. Taking now into
account the finite support of the wave packet $h(x)$ whose size is supposed to be much smaller than the
characteristic curvature scale $l_c$, we obtain
\beqa\nonumber
\hat{a}^\dagger(h) &=& i\int_\sigma d^3\mathbf{y}
\Big(\hat{\Phi}^\dagger(y)\,\partial_{y^0}h(y) - h(y)\,\partial_{y^0}\hat{\Phi}^\dagger(y)\Big) \quad
\text{with} \quad y_0 \;=\; \tau\,.
\eeqa
This is consistent with the Minkowski-space approximation one has been successfully employing in 
particle physics. The physical particle is thus given by
\beqa
|\psi\rangle &=& \hat{a}^\dagger(h)|\Omega\rangle\,,
\eeqa
which is localised over the support of the wave packet $h(x)$ and, hence, is normalisable.

Thus, the equivalence principle plays an essential role in our proposal for defining the physical notion 
of particle at any space-time point of the universe, where the local curvature length is
much larger than the particle size. Since these particles when considered in the local
Minkowski frame are identical to Wigner's ones, we can employ the standard methods of the Feynman
rules and diagrams to describe their scattering reactions.

\section{Self-interacting scalar field in Schwarzschild space}
\label{sec:scalar-model}

We shall consider a massless scalar field with the conformal coupling to gravity and the quartic 
self-interacting term in the background of an evaporating and spherically symmetric black hole.
Specifically, the Lagrangian $\mathcal{L}$ is given by
\beqa\label{eq:lofsf1}
\mathcal{L} &=& -\frac{1}{2}\,\Phi{\Box}\Phi + \frac{1}{12}R\,\Phi^2 - \frac{\lambda}{4!}\,\Phi^4\,,
\eeqa
where $R$ is the Ricci scalar. The Ricci scalar vanishes in the Schwarzschild geometry which is described 
by the line element
\beqa\label{eq:line-element}
ds^2 &=& g_{\mu\nu}dx^\mu dx^\nu \;=\; f(r)dt^2 - \frac{dr^2}{f(r)} - r^2d\Omega^2\,, \quad
\text{where} \quad f(r) \;=\; 1 - \frac{r_H}{r}\,,
\eeqa
where $r_H = 2M$ is a size of the black-hole horizon of mass $M$. In the following we shall use
the surface gravity $\kappa$ on the horizon which is defined as $\kappa \equiv
\frac{1}{2}f'(r_H) = 1/2r_H$.

A freely-falling frame in Schwarzschild space is characterised by an affine parameter $\tau$. This
parameter corresponds to the Painlev\'{e}-Gullstrand time (see, e.g.,~\cite{Martel&Poisson} for a 
brief review). The line element~\eqref{eq:line-element} in the Painlev\'{e}-Gullstrand coordinates reads
\beqa
ds^2 &=& f(r)d\tau^2 - 2\big(1 - f (r)\big)^{\frac{1}{2}}d\tau dr - dr^2 - r^2d\Omega^2\,.
\eeqa
The time coordinate $\tau$ reduces to the standard Minkowski (M) time $t_M$ for $r \gg r_H$. That is also the case
for the Schwarzschild (S) time $t_{S}$ for $r \rightarrow \infty$.\footnote{Note that a
stationary observer, i.e. the observer moving along the Killing vector $\partial_{t_S}$, possesses
a non-trivial acceleration, which asymptotically vanishes in the spatial infinity.} However, the 
Painlev\'{e}-Gullstrand and Schwarzschild time considerably differ from each other in the near-horizon 
region. We shall establish the relation between these times at $r \sim r_H$ in what follows.

\subsection{Wightman function}

We found in~\cite{Emelyanov-2016-2} the Wightman two-point function
of the non-interacting ($\lambda = 0$), massless scalar field for the Unruh (U) state~\cite{Unruh}. 
In this paper, we want to approximate the local Minkowski vacuum $|\Omega\rangle$ 
by the Unruh state in the far- and near-horizon region. This approximation is sufficiently accurate 
at $r \gg r_H$. In case of $r \sim r_H$, the approximation is still adequate up to some corrections (we
shall come back to this issue below).

The 2-point function in the Unruh state reads
\beqa\label{eq:unruh-2-point-function}
W_U(x,x') &=& \vec{W}_{\beta}(x,x') + \cev{W}(x,x') 
\\[1mm]\nonumber
&=&
\int\limits_0^{+\infty}d\omega
\left(\frac{\cos\left(\omega\Delta{t} + 
i\frac{\omega\beta}{2}\right)}{4\pi\omega\sinh\left(\frac{\beta\omega}{2}\right)}\,\vec{K}_\omega({\bf{x}},{{\bf{x}}}') + 
\frac{\exp\left(-i\omega\Delta{t}\right)}{4\pi\omega}\,\cev{K}_\omega({\bf{x}},{{\bf{x}}}')\right),
\eeqa
where $\beta \equiv 1/T_H$ is the inverse Hawking temperature. The right arrow refers to the ``outgoing" 
modes, whereas the left one to the ``ingoing" modes. The functions
$\vec{K}_\omega({\bf{x}},{{\bf{x}}}')$ and $\cev{K}_\omega({\bf{x}},{{\bf{x}}}')$ are given by
\beqa\label{eq:kvec}
\vec{K}_\omega({\bf{x}},{{\bf{x}}}') &\approx& 
\frac{\bar{\Delta}^{\frac{1}{2}}(\rho)\sin(\omega\rho)}{4\pi\omega\rho(f(r)f(r'))^{\frac{1}{2}}}
\left\{
\begin{array}{ll}
4\omega^2 - \frac{(f(r)f(r'))^{\frac{1}{2}}}{rr'}\,\Gamma_\omega\,, & r \sim r_H\,, \\[3mm]
\frac{(f(r)f(r'))^{\frac{1}{2}}}{rr'}\,\Gamma_\omega\,, & r \gg r_H\,,
\end{array}
\right.
\eeqa
and
\beqa\label{eq:kcev}
\cev{K}_\omega({\bf{x}},{{\bf{x}}}') &\approx& 
\frac{\bar{\Delta}^{\frac{1}{2}}(\rho)\sin(\omega\rho)}{4\pi\omega\rho(f(r)f(r'))^{\frac{1}{2}}}
\left\{
\begin{array}{ll}
\frac{(f(r)f(r'))^{\frac{1}{2}}}{rr'}\,\Gamma_\omega\,, & r \sim r_H\,, \\[3mm]
4\omega^2 - \frac{(f(r)f(r'))^{\frac{1}{2}}}{rr'}\,\Gamma_\omega\,, & r \gg r_H\,,
\end{array}
\right.
\eeqa
where $\rho \equiv (2\bar{\sigma}({\bf{x}},{{\bf{x}}}'))^{\frac{1}{2}}$, $\bar{\sigma}({\bf{x}},{{\bf{x}}}')$ 
is the three-dimensional geodetic interval for the ultra-static or optical metric 
$\bar{g}_{\mu\nu} = g_{\mu\nu}/f(r)$, $\bar{\Delta}(x,x')$ is the Van Vleck-Morette
determinant and 
\beqa\label{eq:gamma}
\Gamma_\omega &\equiv& \sum_{l = 0}^{+\infty}(2l + 1)|B_{\omega l}|^2 \;\approx\; 27\omega^2M^2
\eeqa
in the DeWitt approximation~\cite{DeWitt}.

The origin of the ``ingoing" and ``outgoing" part of the Wightman function $W_U(x,x')$ can be understood 
as follows. The scalar field operator $\hat{\Phi}(x)$ is represented through a sum of the non-Hermitian 
operators, namely $\hat{\Phi}(x) = \hat{a}(x) + \hat{a}^\dagger(x)$, where $\hat{a}(x)$ annihilates the vacuum, 
i.e. $\hat{a}(x)|\Omega\rangle = 0$ (cf. Eq.~\eqref{eq:field-splitting}). This operator can in turn be 
represented as
\beqa
\hat{a}(x) &=& \hat{a}_>(x) + \hat{a}_<(x)\,,
\eeqa
such that $\hat{a}_<(x)|\Omega\rangle = \hat{a}_>(x)|\Omega\rangle = 0$, but $\hat{a}_<(x)$
has a non-vanishing support only for the advanced Finkelstein-Eddington time $v < v_H$, while 
$\hat{a}_>(x)$ possesses a non-vanishing support for $v > v_H$, where $v_H$ corresponds to 
the moment when the event horizon has formed (see, e.g.,~\cite{DeWitt}).\footnote{In 
Minkowski space this splitting
can be done, e.g., with respect to the origin of the reference frame, such that $\hat{a}_>(x)$ and 
$\hat{a}_<(x)$ have a support only on $\Sigma_>$ and $\Sigma_<$, respectively, where 
$\Sigma_>$ is a part of the Cauchy surface for $x >0$, whereas $\Sigma_<$ for $x < 0$.
The total Cauchy surface $\Sigma$ in space is thus given by $\Sigma_<{\cup}\Sigma_>$.}
Since the operators $\hat{a}_<(x)$ and $\hat{a}_>(x)$ have non-intersecting supports, it generally 
holds
\beqa
[\hat{a}_<(x),\hat{a}_>(x')] &=& [\hat{a}_<(x),\hat{a}_>^\dagger(x')] \;=\; 0\,.
\eeqa
The operator $\hat{a}_<(x) + \text{H.c.}$ is further split as follows~\cite{Hawking}:
\beqa
\hat{a}_<(x) + \text{H.c.} &=& \hat{b}(x) + \hat{c}(x) + \text{H.c.}\,,
\eeqa 
where $\hat{b}(x)$ and $\hat{c}(x)$ have a non-vanishing support above and under the
horizon, respectively, and $\hat{b}(x)|\tilde{\Omega}\rangle = \hat{c}(x)|\tilde{\Omega}\rangle = 0$,
where $|\tilde{\Omega}\rangle$ is the Boulware vacuum~\cite{Boulware}. These operators commute
with each other as possessing non-intersecting supports:
\beqa
[\hat{b}(x),\hat{c}(x')] &=& [\hat{b}(x),\hat{c}^\dagger(x')] \;=\; 0\,.
\eeqa
The scalar field operator expressed through these operators becomes
\beqa\label{eq:scala-operator-splitting}
\hat{\Phi}(x) &=& \hat{\Phi}_>(x) + \hat{\Phi}_<(x) \;=\;
\hat{\Phi}_>(x) + \hat{\Phi}_b(x) + \hat{\Phi}_c(x)\,,
\eeqa
and, hence, the Wightman function $W_U(x,x')$ above the horizon ($\hat{\Phi}_c(x) = 0$) is
\beqa\nonumber
\langle\hat{\Phi}(x)\hat{\Phi}(x')\rangle &=& 
\langle\hat{\Phi}_>(x)\hat{\Phi}_>(x')\rangle + 
\langle\hat{\Phi}_b(x)\hat{\Phi}_b(x')\rangle + \langle\hat{\Phi}_c(x)\hat{\Phi}_c(x')\rangle
\\[2mm]
&=& \langle\hat{\Phi}_>(x)\hat{\Phi}_>(x')\rangle + 
\langle\hat{\Phi}_b(x)\hat{\Phi}_b(x')\rangle \;=\; \cev{W}(x,x') + \vec{W}_{\beta}(x,x')\,.
\eeqa
Therefore, the ``ingoing" part of the 2-point function is due to the operator $\hat{\Phi}_>(x)$, while the ``outgoing"
one originates from the operator $\hat{\Phi}_b(x)$. Note that the vacuum $|\Omega\rangle$ responds
to the action of the operator $\hat{\Phi}_b(x)$ as a thermal state at the Hawking temperature $T_H$ defined 
with respect to the Schwarzschild time $t_S$, while as an empty state when probed by $\hat{\Phi}_>(x)$.
In general, any polynomial composed of the operator $\hat{\Phi}_>(x)$ probes $|\Omega\rangle$ as
an empty state, while composed of $\hat{\Phi}_b(x)$ as if $|\Omega\rangle$ is a mixed state. It is worth
emphasising that the latter effect is due to the field operator $\hat{\Phi}_b(x)$, rather than the vacuum 
state (see for an earlier version of this point~\cite{Emelyanov-2014-2,Emelyanov-2015-1}).\footnote{This 
can be elucidated as follows. Consider a quantum-mechanical system of two 
non-interacting harmonic oscillators $\{\hat{a},\hat{a}^\dagger\}$ and $\{\hat{b},\hat{b}^\dagger\}$ of the
same frequency $\omega$ and with a ground state $|0\rangle = |0_a\rangle{\otimes}|0_b\rangle$, such
that $\langle 0|\hat{a}^\dagger\hat{a}|0\rangle = \langle 0|\hat{b}^\dagger\hat{b}|0\rangle = 0$. Perform
now a double squeezed transformation, i.e. $\hat{\alpha} = \cosh\theta\,\hat{a} - \sinh\theta\,\hat{b}^\dagger$
and $\hat{\beta} = \cosh\theta\,\hat{b} - \sinh\theta\,\hat{a}^\dagger$. It is straightforward to show that
$[\hat{\alpha},\hat{\beta}] = [\hat{\alpha},\hat{\beta}^\dagger] = 0$, i.e. the oscillators 
$\{\hat{\alpha},\hat{\alpha}^\dagger\}$ and $\{\hat{\beta},\hat{\beta}^\dagger\}$ are independent.
If one chooses the parameter $\theta$ of the transformation to satisfy 
$\tanh\theta = \exp(-\omega\beta/2)$, then one has, e.g., 
$\langle 0|\hat{\alpha}^\dagger\hat{\alpha}|0\rangle = 1/(\exp(\beta\omega) - 1)
= \text{tr}(\hat{\rho}_\beta\hat{\alpha}^\dagger\hat{\alpha})$, where $\hat{\rho}_\beta$ is a density matrix
of inverse temperature $\beta$. Thus, one may say that the ground state $|0\rangle$ is
a mixed state when probed by operators of the type $\hat{\alpha}$. Thus, a pure state
can sometimes respond as a mixed state due to the non-triviality of quantum operators. Note 
that a ground state $|\tilde{0}\rangle$ annihilated by both $\hat{\alpha}$ and $\hat{\beta}$
(but not by $\hat{a}$ or $\hat{b}$) can be mapped to $|0\rangle$ by a unitary operator. It is an easy
exercise to show that and is actually guaranteed by the Stone-von Neumann theorem. This is not
anymore the case in quantum field theory (QFT), where this map is \emph{not} unitarily implementable. 
This fact allows in particular to describe phase transitions (e.g. normal phase $\leftrightarrow$ superconductive 
phase) in the framework of 
QFT, which is impossible in quantum mechanics. Note that, for that reason, toy models in the background 
of black holes based on qubits presuppose a physical realisation of the Hawking excitations. Therefore,
this kind of the models cannot anyhow provide a resolution of the information loss problem, but can and does
cause further confusions.}

It is tempting to conclude that particles $|\tilde{\psi}\rangle$ defined as 
\beqa
|\tilde\psi\rangle &=& \hat{b}^\dagger(\tilde{h})|\tilde{\Omega}\rangle
\eeqa
are thermally populated in the vacuum $|\Omega\rangle$, where $\tilde{h}(x)$ is a wave packet being 
a positive frequency solution of the field equation with respect to $t_S$ with definite
values of the orbital and magnetic numbers. However, 
one can show that $\langle\tilde{\psi}|\Omega\rangle$ is identically zero, i.e.
$\langle\tilde{\psi}|\Omega\rangle = 0$. Moreover, $\langle\tilde{\Omega}|\Omega\rangle = 0$ and, 
hence, the vacua $|\Omega\rangle$ and $|\tilde{\Omega}\rangle$ give unitarily inequivalent Hilbert 
space representations ($\mathcal{H}$ and $\tilde{\mathcal{H}}$, respectively) of the same operator 
algebra $\mathcal{A}$~\cite{Emelyanov-2015-2}. This means that the physical interpretation of the
state $|\Omega\rangle$ as a thermal state of the particles $|\tilde\psi\rangle$ at the Hawking 
temperature $T_H$ is not self-consistent, although most of the researchers take the contrary for 
granted. We shall come back to this issue below.

\subsection{One-loop correction to self-energy}

In general, it is hardly possible to obtain the full propagator $G(x,x')$ in the non-linear theories.
This exact propagator in the massless $\lambda\Phi^4$-theory up to the 1-loop order satisfies 
\beqa
\Big(\Box + m_\Phi^2 + \text{O}\big(\lambda^2\big)\Big)G(x,x') &=& \frac{-i}{(-g(x))^\frac{1}{2}}\,\delta\big(x-x'\big)\,,
\eeqa
Expanding the propagator $G(x,x')$ through the free propagator $G_U(x,x')$ by employing the standard 
Feynman rules for this theory, one can obtain the effective mass of the scalar field at one-loop approximation.
This can be pictorially expressed as follows
\beqa
m_\Phi^2\,G_U(x,x') &=& -\Box_x\left(\mathbf{\imineq{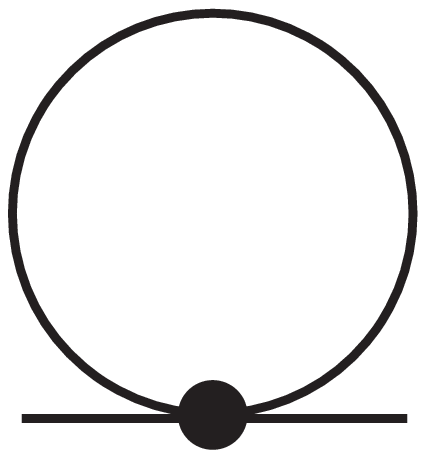}{5.5}}
\right).
\eeqa

To compute the effective mass of the scalar field, one thus needs to establish the Feynman propagator
$G_U(x,x')$ in the Unruh state that approximates the local Minkowski one $|\Omega\rangle$.
This vacuum state is supposed to be physically (unitarily) equivalent to the vacuum state before the black
hole has formed. In other words, we do not expect the change of quantum physics (the change of the
Hilbert space representation of the field operator algebra or the phase transition) as a result of the
black-hole formation.

\subsubsection{Far-horizon region: $R \gg r_H$}
\label{subsubsec:weak}
Far away from the black hole, i.e. $R \gg r_H$, where $R$ is the distance to the centre of the black hole, one
can approximate geometry by Minkowski space. It implies that the geodetic interval for the optical and 
physical metric approximately coincide and read
\beqa\label{eq:tadpole-far}
2\sigma(x,x') &\approx&
2\bar{\sigma}(x,x') \;\approx\; (t-t')^2 - (\mathbf{x}-\mathbf{x}')^2\,,
\eeqa
where $\mathbf{x} = (r\cos\theta\cos\phi,r\cos\theta\sin\phi,r\sin\theta)$. It should be noted
that the Schwarzschild time $t_S$ approaches the Painlev\'{e}-Gullstrand time $\tau$ in the asymptotically flat
region as 
\beqa
t_S &=& \tau\,\big(1 + \text{O}(r_H/R)\big)\,, 
\eeqa
such that $t_S \rightarrow \tau$ for $R \rightarrow \infty$.

The 2-point function in the asymptotically flat region is
\beqa\label{eq:correlator-far}
W_U(x,x') &\approx& \Big(1 - \frac{27r_H^2}{16R^2}\Big) W_M(x,x') + \frac{27r_H^2}{16R^2}\,W_M^\beta(x,x')\,
\eeqa
for $|x-x'| \ll R$, where $W_M^\beta(x,x')$ is functionally given by the Minkowski two-point function at the 
inverse temperature $\beta = 1/T_H$, which becomes $W_M(x,x')$ for $T_H \rightarrow 0$. In the limit 
$R \rightarrow \infty$, $W_U(x,x')$ reduces to the standard Minkowski correlator $W_M(x,x')$ and, hence, 
the influence of the black hole on local physics can be fully neglected.\footnote{It is not the case for eternal
black holes. This does not serve a problem, because black holes of this type are not realistic. However, if
a tiny black hole of this type could appear as a result of a quantum space-time fluctuation, then its influence
on local physics cannot be neglected even at spatial infinity. For instance, photons acquire
an effective thermal mass of the order of $\alpha^\frac{1}{2}T_H$ for $T_H \gg m_e$ in
QED, assuming such a black hole exists for a sufficiently large time interval~\cite{Emelyanov-2016-2}.
These are certainly ruled out if the vacuum state is given by the Hartle-Hawking one.}

The Feynman propagator $G_U(x,x')$ can be expressed through the commutator function and reads
\beqa\label{eq:propagator-far}
G_U(k,k') &\approx& \left(\frac{i}{k^2 + i\varepsilon} 
+ 2\pi\left(\frac{27r_H^2}{16R^2}\right)\frac{\delta\big(k^2\big)}{e^{|k_0|/T_H} - 1}\right)\delta\big(k-k'\big)
\eeqa
in the momentum representation~\cite{Emelyanov-2016-2}. The frequency $k_0$ here is defined
with respect to the Painlev\'{e}-Gullstrand time $\tau$. The second term on the right-hand side
of~\eqref{eq:propagator-far} depends on the distance to the black hole and vanishes in the limit 
$R \rightarrow \infty$. Again, it means that local physics does not change in the asymptotic region, 
where the field propagator is insensitive to the black-hole properties.

Thus, one obtains
\beqa\label{eq:smfar}
\Box_x\left(\mathbf{\imineq{1-loop-self-phi4.eps}{5.5}}\right) &=& 
-\lambda\int dx_1(-g(x_1))^{\frac{1}{2}}\;\Box_xG_U(x,x_1)G_U(x_1,x_1)G_U(x_1,x')
\\[2mm]\nonumber
&=& -\lambda G_U(x,x)G_U(x,x')
\;\approx\; -\lambda\int\frac{d^3\mathbf{k}}{(2\pi)^3}\frac{1}{|\mathbf{k}|}
\left(\frac{1}{2} + \frac{27r_H^2}{16R^2}\frac{1}{e^{|\mathbf{k}|/T_H} - 1}\right)G_U(x,x')\,.
\eeqa
The integral over $\mathbf{k}$ in~\eqref{eq:smfar} diverges unless one subtracts the first term in
the parenthesis. In Minkowski space with a hot physical plasma, one can renormalise this by adding 
a mass counter-term to the Lagrangian density. This essentially implies that one
removes this UV divergence by subtracting all terms which do
not vanish in the limit of the vanishing plasma  temperature. In the black-hole background, this can
be accounted for the divergent Boulware contribution to the scalar mass.

Having got rid of the ultraviolet divergence in~\eqref{eq:smfar}, we obtain
\beqa\label{eq:smfar-reg}
\Box_x\left(\mathbf{\imineq{1-loop-self-phi4.eps}{5.5}}\right) &\approx& 
-\frac{\lambda\,\xi}{16\pi^2R^2}\,G(x,x') \quad
\text{with} \quad \xi \;\equiv\; \frac{9}{128}\,.
\eeqa
The effective mass of the scalar field is thus finite and reads
\beqa
m_\Phi^2 &\approx& \frac{\lambda\,\xi}{16\pi^2R^2}\,.
\eeqa
The scalar-field mass $m_\Phi$ turns out not to depend on the black-hole mass $M$ in the
leading order of the approximation. It appears to be the case, because \eqref{eq:smfar-reg}
equals $T_H(r_H/R) \propto 1/R$ up to a numerical factor. We found a similar property of the one-loop
correction to the photon self-energy in the case of evaporating black holes of mass $M \ll 10^{16}\,\text{g}$
in~\cite{Emelyanov-2016-2}.

\subsubsection{Near-horizon region: $R \sim r_H$}
\label{sec:olctse-nhr}
The Wightman function $W_U(x,x')$ has a geometrical prefactor $1/f(r)$. Therefore, it is
tempting to conclude that the 
loop diagrams are divergent on the horizon. For instance, the 1-loop correction to the self-energy of the 
scalar field seems to increase as $1/f(r)$ in the near-horizon region, while as $1/f^2(r)$ at 2-loop level.
However, this does not happen to be the case if one studies this carefully in the freely-falling frame.

To analyse near-horizon physics, we introduce a local Minkowski frame at $R \sim r_H$ of a 
black hole of mass $M \gtrsim M_\odot$. The size of the local Minkowski frame is then about
\beqa
l_M &\gtrsim& 300{\times}(M/M_\odot)\,\text{m}\,.
\eeqa 
The local dynamics in this frame is set by the geodesic vector $G$, such that $G_\mu = \{1,0,0,0\}$
in the Painlev\'{e}-Gullstrand coordinates with the time coordinate $\tau$. This vector $G$ is one of the
generators of the local Poincar\'{e} group whose irreducible representations correspond to the
particle states.

The light-cone Kruskal-Szekeres coordinates $U$ and $V$ near the horizon behave as
\bsubeqs\label{eq:kruskal-coordinates}
\beqa
U &=& \alpha\Big(\tau - \tau_0 - 2\kappa(\tau - \tau_0)^2 + \text{O}\big((\tau - \tau_0)^3\big)\Big)\,,
\\[1mm]
V &=& (e/\alpha)\Big(2/\kappa  + \tau - \tau_0 + \text{O}\big((\tau - \tau_0)^3\big)\Big)\,,
\eeqa
\esubeqs
where $e$ is the Euler number and $0 < \alpha \leq e^\frac{1}{2}$. The event horizon corresponds to 
$U_H = 0$ or $\tau = \tau_0 > 0$, while $V_H = 2e/(\alpha\kappa)$ holds at $\tau_0$. This reveals the
geometrical meaning of the constant $\alpha$ in~\eqref{eq:kruskal-coordinates}.
Introducing $\tau' = (V+U)/2e^{\frac{1}{2}}$ and $x' = (V-U)/2e^{\frac{1}{2}}$, we obtain
\beqa
ds^2 &\approx& d\tau'{}^2 - dx'{}^2 - dy^2 - dz^2\,
\eeqa
near the horizon, where $y^2+z^2 = 4r_H^2\tan^2(\theta/2)$, $z/y = \tan\phi$ and $y^2 + z^2 \ll r_H^2$. 
Employing the local Lorentz transformation with $v/c = (e-\alpha^2)/(e + \alpha^2)$, the time coordinate 
$\tau'$ can be transformed to the Painlev\'{e}-Gullstrand time $\tau$ (up to a translation).

On the other hand, the Schwarzschild metric in the near-horizon region can be approximated by the 
Rindler metric. Specifically, it holds
\beqa
ds^2 &\approx& \kappa^2\rho^2dt^2 - d\rho^2 - dy^2-dz^2\,,
\eeqa
where $\rho = \int dr/f^{\frac{1}{2}}(r) \approx (4r_H(r-r_H))^{\frac{1}{2}}$. This line element 
can be further transformed into Minkowski one via the diffeomorphism 
$\tau = \rho\sinh(\kappa t)$ and $x = \rho\cosh(\kappa t)$:
\beqa
ds^2 &\approx& d\tau^2 - dx^2 - dy^2 - dz^2\,.
\eeqa
Note that we have tacitly introduced a new time $\tau$ which is up to the Lorentz transformation
coincides with the Painlev\'{e}-Gullstrand time. Therefore, we have denoted both by the same symbol.
  
We now want to study local physics in this coordinate system. For this
purpose we need to derive the Feynman propagator $G_U(x,x')$ in the freely-falling frame. For the
reasons which become clear later on, we study first the vacuum expectation value of the operator
$\hat{\Phi}^2(x)$.

\subparagraph{Wick squared operator $\hat{\Phi}^2(x)$ near horizon}

Using the background-field method to compute the 1-loop contribution to the scalar field equation, 
we find
\beqa
\Big(\Box + \frac{\lambda}{2}\,\langle\hat{\Phi}^2(x)\rangle\Big)\Phi(x) 
&=& 0\,,
\eeqa
where $\langle\hat{\Phi}^2(x)\rangle$ has been appropriately renormalised (see below). Thus, the effective 
scalar mass at 1-loop level is given by
\beqa\label{eq:smws}
m_\Phi^2 &=& \frac{\lambda}{2}\,\langle\hat{\Phi}^2(x)\rangle\,.
\eeqa

The renormalised value of the Wick squared operator $\hat{\Phi}^2(x)$ in the Unruh vacuum
was computed in~\cite{Candelas}. This quantity turns out to be finite on the horizon and decreases as $1/R^2$
at the spatial infinity. Specifically, it holds that
\beqa\label{eq:wick-squared-operator}
\langle \hat{\Phi}^2(x)\rangle &\approx&
\left\{
\begin{array}{ll}
\big(\frac{1}{3} - 2\xi\big)T_H^2\,, & R \sim r_H\,, \\[3mm]
\xi/8\pi^2R^2\,, & R \gg r_H\,.
\end{array}
\right.
\eeqa
This is in full agreement with our result obtained above for $R \gg r_H$ using the diagrammatic approach.

The finiteness of $\langle \hat{\Phi}^2(x)\rangle$ on the black-hole horizon seems \emph{a priori} not to be 
guaranteed. Indeed, the Wick squared operator is defined in general as
\beqa
\hat{\Phi}^2(x) &=& \lim\limits_{x' \rightarrow x}\left(\hat{\Phi}(x)\hat{\Phi}(x') - H(x,x')\hat{1}\right),
\eeqa
where $H(x,x')$ is the Hadamard parametrix. It is a geometrical (state-independent) object and designed 
to subtract the ultraviolet divergences in the 2-point function only. In our case, the Hadamard parametrix
is of the form
\beqa
H(x,x') &=& -\frac{1}{8\pi^2\sigma(x,x')} + \textrm{O}\big(\sigma\ln\sigma\big)\,,
\eeqa
where $\sigma(x,x')$ is the geodetic interval for the physical metric. It is worth mentioning that the term
$\sigma\ln\sigma$ in $H(x,x')$ is fully responsible for the trace aka conformal
anomaly~\cite{Moretti,Decanini&Folacci}. 

The parametrix can be expressed via the geodetic interval in the optical metric:
\beqa\label{eq:hp}
H(x,x') &=& -\frac{\left(f(r)f(r')\right)^{-\frac{1}{2}}}{8\pi^2\bar{\sigma}(x,x')}
+ \frac{M^2f^{-1}(r)}{48\pi^2r^4} + \textrm{O}\big(\bar{\sigma}\ln\bar{\sigma}\big)\,.
\eeqa
The first term in~\eqref{eq:hp} coincides with the 2-point function $W_B(x,x')$ in the
Boulware vacuum for $x \sim x'$, whereas the second term in~\eqref{eq:hp} gives a non-vanishing
value of $-\langle\tilde{\Omega}| \hat{\Phi}^2(x)|\tilde{\Omega}\rangle$ far away
from the event horizon which disappears in the limit $M \rightarrow 0$~\cite{Candelas}.

In the far-from-horizon region, the geodetic intervals $\sigma(x,x')$ and $\bar{\sigma}(x,x')$
go over to the Minkowski geodetic distance. These significantly differ from each other in the near-horizon
region. We obtain that
\beqa\label{eq:gdnh}
\sigma(x,x') &=& \sigma_0(x,x') 
\\[1mm]\nonumber
&& + \;\frac{1}{24} \left(\frac{f'(r)^2-4\kappa^2}{f(r)}\frac{f'(r')^2-4\kappa^2}{f(r')}\right)^{\frac{1}{2}}\sigma_0^2(x,x') 
+ \textrm{O}\big(\sigma_0^3(x,x')\big)\,,
\eeqa
where we have taken into account that $f'(r_H) = 1/r_H = 2\kappa$ and
\beqa
2\sigma_0(x,x') &\approx& \Delta\tau^2 - \Delta{x}^2 - \Delta{y}^2 - \Delta{z}^2
\eeqa 
close to the horizon, while $\bar{\sigma}(x,x')$ can be found in~\cite{Emelyanov-2015-1} and approaches 
in the limit $R \rightarrow r_H$ to the geodetic interval of the static space with the hyperbolic spatial 
section. Substituting $\sigma(x,x')$ into the Hadamard parametrix, we find that
\beqa\label{eq:hadamard-parametrix-lif-near-horizon}
H(x,x') &=& -\frac{1}{8\pi^2\sigma_0(x,x')} - \frac{\kappa^2}{12\pi^2} 
+ \textrm{O}\big(\sigma_0\ln\sigma_0,\kappa^2f(r)\big)\,.
\eeqa

This result and \eqref{eq:kvec} with \eqref{eq:kcev} at $R \sim r_H$ in the local Minkowski
frame in turn allow us to compute the Wightman function in the near-horizon region:
\beqa\label{eq:2-point-function-neat-horizon}
W_U(x,x') &\approx& -\frac{1}{8\pi^2\sigma_0(x,x')} - \frac{\xi\kappa^2}{2\pi^2}\,.
\eeqa
Thus, we reproduce the result \eqref{eq:wick-squared-operator} obtained in~\cite{Candelas} by subtracting 
$H(x,x)$ from $W_U(x,x)$.\footnote{It is worth noticing that the finite terms in $W_U(x,x')$ and $H(x,x')$ in 
the coincidence limit $x' \rightarrow x$ are separately regular on the event horizon $r = r_H$ in the freely-falling
frame. In the Schwarzschild frame, these parts of the Wightman function and the Hadamard
parametrix increase as $1/f(r)$ for $r \rightarrow r_H$, but their difference turns out to be non-singular at 
$r = r_H$ as found in~\cite{Candelas}.}

\subparagraph{Feynman propagator near horizon}

The Feynman propagator can be expressed through the commutator function $C(x,x')$ which is equal to
$W_U(x,x') - W_U(x',x)$. The commutator $C(x,x')$ is insensitive to the time-independent term in the
correlation function. There are at least two possibilities to deal with the second term in $W_U(x,x')$:
Either one needs to 
introduce an imaginary ``temperature" $\Theta \sim i\kappa$ or a discrete frequency spectrum 
$\omega_n \sim \kappa n$. Since we expect merely a slight change of physics in the local inertial frame 
even near the horizon of a large black hole, we do not consider the possibility of the
change of the continuous 
spectrum into the discrete one.\footnote{It is worth mentioning that the finite term in the coincidence limit 
$x' \rightarrow x$ of the thermal two-point function of temperature $T$ is given by $+T^2/12$. However, 
the Wightman function in the near-horizon region given in~\eqref{eq:2-point-function-neat-horizon} has 
a negative correction to the term $-1/\sigma_0(x,x')$. This effectively corresponds to the imaginary 
``temperature".}

The time-independent term in \eqref{eq:2-point-function-neat-horizon} is negligibly small, because we have
been working in the regime $\sigma_0\kappa^2 \ll 1$. To reproduce the value of 
$\langle \hat{\Phi}^2(x)\rangle$ through the tadpole diagram, we thus define
\beqa\label{eq:propagator-near}
G_U(p,p') &\approx& \left(\frac{i}{p^2 + i\varepsilon} + \frac{2\pi\delta(p^2)}{e^{|p_0|/\Theta_\varepsilon} - 1}\right)
\delta\big(p-p'\big)\,,
\eeqa
where we have introduced an \emph{effective} imaginary ``temperature":
\beqa\label{eq:imaginary-temperature}
\Theta_\varepsilon &\equiv& \varepsilon + \frac{i\kappa}{\pi}\,(6\xi)^{\frac{1}{2}} \quad \text{with} \quad
\varepsilon \;\rightarrow\; +0\,.
\eeqa
Since $\langle \hat{\Phi}^2(x)\rangle \propto G_U(x,x)$ from the tadpole diagram 
(see Eq.~\eqref{eq:smfar}), one needs to renormalise it by subtracting the ``Hadamard propagator" at $x=x'$.
We define it as follows
\beqa\label{eq:propagator-hadamard}
G_H(p,p') &\approx& \left(\frac{i}{p^2 + i\varepsilon} + \frac{2\pi\delta(p^2)}{e^{|p_0|/\theta_\varepsilon} - 1}\right)
\delta\big(p-p'\big)\,, \quad \textrm{where} \quad 
\theta_\varepsilon \;\equiv\; \varepsilon + \frac{i\kappa}{\pi}\,.
\eeqa

We can now reproduce the result of~\cite{Candelas} for the Wick squared or the effective scalar mass
if we renormalise the ultraviolet divergence of the tadpole diagram by subtracting $G_H(x,x')$ from $G_U(x,x')$.
Specifically, we have
\beqa\label{eq:effective-mass-diagram}
m_\Phi^2 &=& \frac{\lambda}{2}\Big(G_U(x,x) - G_H(x,x)\Big)
\\[2mm]\nonumber
&\approx& \frac{\lambda}{2}\int\frac{d^3\mathbf{k}}{(2\pi)^3}\frac{1}{|\mathbf{k}|}
\left(\frac{1}{e^{|\mathbf{k}|/\Theta_\varepsilon} - 1} - \frac{1}{e^{|\mathbf{k}|/\theta_\varepsilon} - 1}\right)
\;=\; \frac{\lambda}{24}\Big(\Theta^2 - \theta^2\Big) 
\;=\; \frac{\lambda}{2}\Big(\frac{1}{3} - 2\xi\Big)T_H^2\,.
\eeqa
Note that both $\Theta^2$ and $\theta^2$ are negative, but the effective scalar mass $m_\Phi^2$
is positive, because the absolute value of $\theta$ is larger than that of $\Theta$.
It should be noted that the first term of the right-hand side in
Eq.~\eqref{eq:effective-mass-diagram} is due to $\theta$ that comes in turn from the Hadamard
parametrix. It is a geometrical object and its contribution to $m_\Phi^2$ is a result of the coordinate
transformation from the local Rindler geometry into the local Minkowski geometry near the event horizon.

\subsection{Local renormalisation scheme}

Above we have found that the effective scalar mass is finite and coincides with~\eqref{eq:smws} if we add 
the minus ``Hadamard loop" to the Feynman loop. One may represent this subtraction pictorially
as follows
\beqa\label{eq:effective-mass-lrs}
m_\Phi^2\,G(x,x') &=& -\Box_x\left(\mathbf{\imineq{1-loop-self-phi4.eps}{5.5}} + 
\mathbf{\imineq{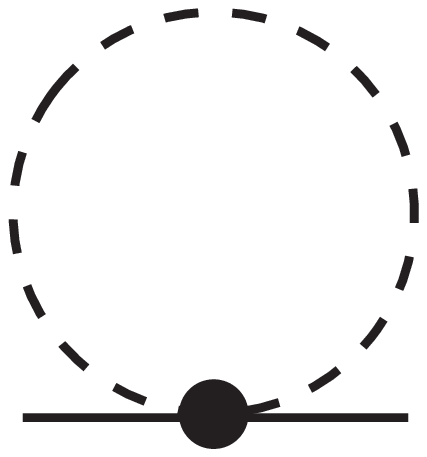}{5.5}} + \text{O}\big(\lambda^2\big)\right),
\eeqa
where the dashed line corresponds to the minus Hadamard propagator, i.e. $-G_H(x,x')$. 

We want to propose a local renormalisation scheme in curved space-times. Specifically, we 
introduce a fictitious scalar field $\phi(x)$ with a \emph{negative} norm (a wrong sign in front of the propagator)
with the propagator being constructed from the Hadamard parametrix. We need also to introduce extra 
vertices in the Lagrangian $\mathcal{L}$, namely
\beqa
\frac{\lambda}{4!}\,\phi^4\,,\quad \frac{4\lambda'}{4!}\,\phi^3\Phi\,,\quad 
\frac{6\lambda}{4!}\,\phi^2\Phi^2\quad \text{and}\quad \frac{4\lambda'}{4!}\,\phi\Phi^3\,,
\eeqa
such those
\bsubeqs
\beqa\label{eq:extra-vertices-1}
\mathbf{\imineq{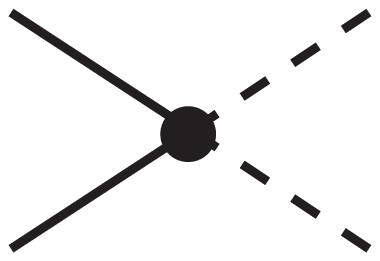}{4.5}} &=&
\mathbf{\imineq{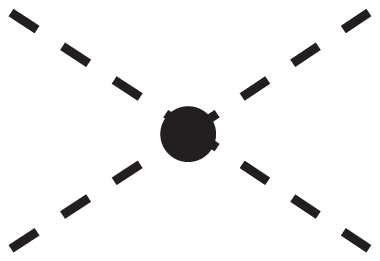}{4.5}} \;=\;-i\lambda\,,
\\[2mm]\label{eq:extra-vertices-2}
\mathbf{\imineq{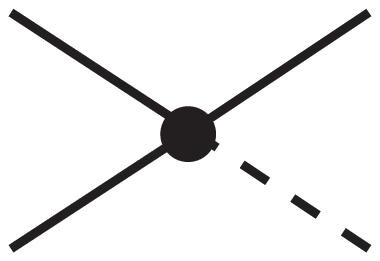}{4.5}} &=& 
\mathbf{\imineq{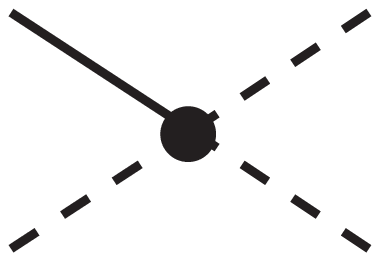}{4.5}} \;=\;-i\lambda'
\eeqa
\esubeqs
in the momentum representation. The ratio $\lambda'/\lambda$ will be fixed below in order to recover
the standard result for the running coupling constant in the asymptotically flat region.

\subsection{One-loop correction to coupling constant}

We now study how the coupling constant $\lambda$ changes in the near-horizon and asymptotically 
flat region at 1-loop level to further investigate the imprints of evaporating black holes in scattering
processes. The four-point
vertex function can be computed by functionally differentiating the path integral 
over the external current which is linearly coupled to the scalar field. The result of this method is by now 
standard and reads
\beqa\label{eq:vertex}
\Gamma^{(4)}(x_i) &=& \mathbf{\imineq{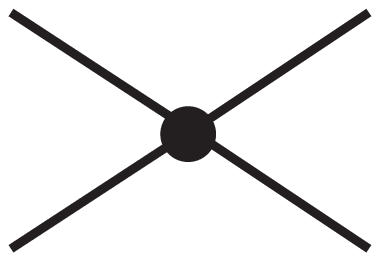}{4.5}}
+ \frac{3}{2}\left(\mathbf{\imineq{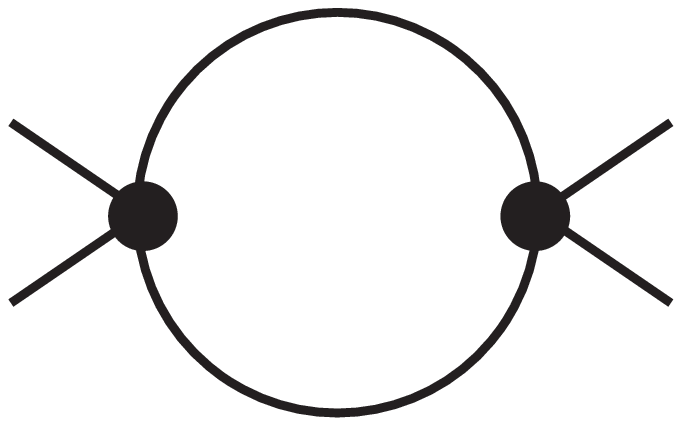}{5.5}}
+2\mathbf{\imineq{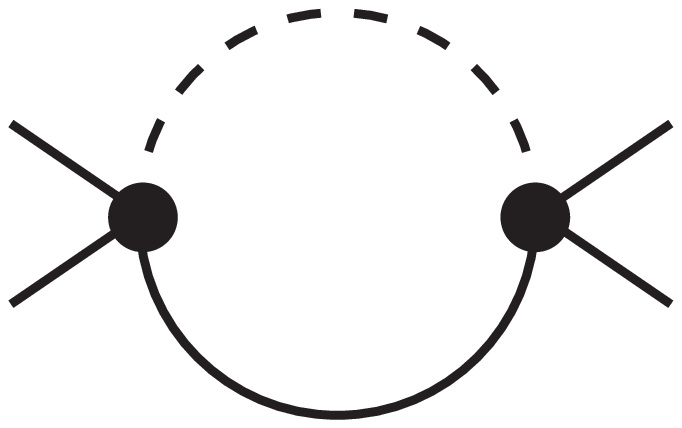}{5.5}}
+\mathbf{\imineq{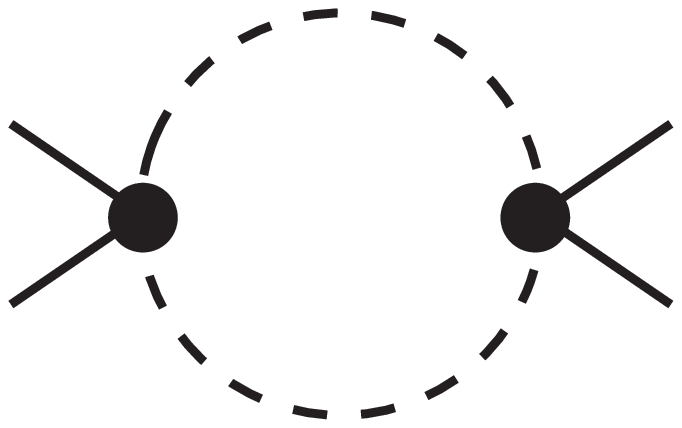}{5.5}}\right) + \text{O}\big(\lambda^3\big)\,,
\eeqa
where $i$ runs from $1$ to $4$. 

\subsubsection{Near-horizon region: $R \sim r_H$}

The first diagram in the momentum representation is given by $-i\lambda$, whereas the first
1-loop diagram in Eq.~\eqref{eq:vertex} is
\beqa\label{eq:2-loop-coupling}
\mathbf{\imineq{1-loop-lambda-phi4.eps}{5.5}} &=& -i\lambda^2
\int\frac{d^4p}{(2\pi)^4}\left(\frac{i}{p^2 + i\varepsilon}\frac{1}{(q-p)^2 + i\varepsilon}
+\frac{4\pi\delta\big(p^2\big)}{e^{|p_0|/\Theta_\varepsilon}-1}\frac{1}{(q-p)^2 + i\varepsilon}\right),
\eeqa
where we have chosen the external momenta be non-exceptional 
($k_0^i = 0$, $\mathbf{k}^i\mathbf{k}^j = q^2(\delta^{ij} - 1/4)$ for the four external legs, i.e. $i,j = 1,2,3,4$
and $q = k_1 + k_2 =- k_3-k_4$)\footnote{Note that the Mandelshtam variables $s = (k_1+k_2)^2$, 
$t = (k_1+k_3)^2$ and $u = (k_1+k_4)^2$ are all equal to $-q^2$ for this choice of the external momenta.
This explains the factor of $3$ in Eq.~\eqref{eq:vertex}.} 
not to have extra IR divergence of the logarithmic type that is due to the zero mass of the scalar 
field~\cite{Kleinert&Frohlinde}. The origin of this divergence is the same as in quantum electrodynamics.
Specifically, the zero photon mass leads to the non-negligible mutual influence of two charged
particles even when these are at infinite distance from each other. This in turn entails the IR divergence
of the $S$-matrix constructed from the asymptotic particle states. 

The first integral in~\eqref{eq:2-loop-coupling} can be evaluated by employing the standard technique
of the dimensional regularisation~\cite{Kleinert&Frohlinde}. The second integral can be simplified after
the integration over the solid angle. The result reads
\beqa
\mathbf{\imineq{1-loop-lambda-phi4.eps}{5.5}} &=& \frac{i\lambda^2}{(4\pi)^2}\left(
\frac{2}{\epsilon} - \gamma + 2 - \ln\Big(\frac{q^2}{4\pi \mu^2}\Big) + 
2\int\limits_0^{+\infty}dx\ln\Big|\frac{1+x}{1-x}\Big| \frac{1}{e^{qx/2\Theta_\varepsilon} -1}\right),
\eeqa
where $\epsilon \rightarrow 0$ in our case, $\gamma$ is the Euler constant and $\mu$ is an arbitrary 
mass scale inherent to the dimensional regularisation. The same result holds for the third 1-loop 
diagram in~\eqref{eq:vertex} after the substitution $\Theta \rightarrow \theta$. 
The second 1-loop diagram in~\eqref{eq:vertex} equals a quarter of the sum of the first and third 
1-loop diagram if we set 
\beqa
\lambda' &=& \lambda/2^{\frac{1}{2}}\,.
\eeqa
Then, after the $\overline{\text{MS}}$ renormalisation, we obtain
\beqa\nonumber\label{eq:lambda-far}
\lambda(q,r_H) &=& \lambda + \frac{3\lambda^2}{32\pi^2}\left(\ln\Big(\frac{q^2}{4\pi \mu^2}\Big) -
\int\limits_0^{+\infty}dx\ln\Big|\frac{1+x}{1-x}\Big| \Big(\frac{1}{e^{qx/2\Theta_\varepsilon} -1} +
\frac{1}{e^{qx/2\theta_\varepsilon} -1}\Big)\right) + \text{O}\big(\lambda^3\big)
\\[2mm]
&=&\lambda + \frac{3\lambda^2}{32\pi^2}\left(\ln\Big(\frac{q^2}{4\pi \mu^2}\Big) 
- \frac{\pi^2}{3q^2}\big(\Theta^2 + \theta^2\big)\right) + \text{O}\big(\lambda^3,\lambda^2\kappa^4/q^4\big)
\eeqa
in the regime $q^2 \gg \kappa^2$ which is consistent with our approximation.

Note that, whenever some loop correction depends on the external momenta, the UV divergence is not
cancelled by the fictitious field. We could rearrange the local renormalisation scheme in a manner that 
the UV divergence of the
diagram \eqref{eq:2-loop-coupling} is absent, but then the entire 1-loop correction to the coupling
constant $\lambda$ would not depend on the external momenta in the limit $\kappa \rightarrow 0$. This 
turns out to be in disagreement 
with the well-known result in particle physics, namely $\lambda$ depends on the energy scale at 
which one is measuring the coupling constant. For this reason, one had to employ the 
dimensional regularisation (or any other standard regularisation) and the
$\overline{\text{MS}}$ renormalisation as well. 

Substituting $\Theta$ and $\theta$ in~Eq.~\eqref{eq:lambda-far}, we find
\beqa\label{eq:1lc-cc-nh}
\lambda(q,r_H) &\approx& \lambda + \frac{3\lambda^2}{32\pi^2}\left(\ln\Big(\frac{q^2}{4\pi \mu^2}\Big) 
+ \Big(\frac{1}{3} + 2\xi\Big)\frac{1}{4r_H^2q^2}\right).
\eeqa
Thus, the coupling constant $\lambda$ becomes stronger at 1-loop level in the near-horizon region.
It should be noted that the correction due to the Schwarzschild black hole is given by
the second term in the parenthesis of Eq.~\eqref{eq:1lc-cc-nh}. This is in turn composed of two
contributions. One of these originates from the Hadamard subtraction, while another (that is
proportional to $\xi$) comes from the correction to the propagator that is related to the black-hole
evaporation.

\subsubsection{Far-horizon region: $R \gg r_H$}

Far away from the horizon, $G_U(x,x')$ is given by~\eqref{eq:propagator-far}. The 
Hadamard parametrix $H(x,x')$ takes the form at the leading order of the approximation
as if there is no black hole, namely
\beqa
H(x,x') &=& -\frac{1}{8\pi^2\sigma_0(x,x')} + \text{O}\big(\sigma_0\ln\sigma_0,M^2/R^4\big)\,.
\eeqa
In this case, we find
\beqa\nonumber
\lambda(q,R) &\approx& \lambda + \frac{3\lambda^2}{32\pi^2}\left(\ln\Big(\frac{q^2}{4\pi \mu^2}\Big) -
\frac{27r_H^2}{16R^2}\int\limits_0^{+\infty}dx\ln\Big|\frac{1+x}{1-x}\Big|\frac{1}{e^{qx/2T_H} -1}\right) 
+ \text{O}\big(\lambda^3\big)
\\[2mm]
&=&\lambda + \frac{3\lambda^2}{32\pi^2}\left(\ln\Big(\frac{q^2}{4\pi \mu^2}\Big) 
- \frac{\xi}{2R^2q^2}\right) + \text{O}\big(\lambda^3,\lambda^2\kappa^2/q^4R^2\big)
\eeqa
Thus, we reproduce the standard result known in particle physics far away ($R \gg r_H$) from
an evaporating black hole. For a large, but fixed $R$, the vacuum polarisation
induced by the black hole slightly suppresses the coupling constant at one-loop approximation.

\section{Local particle physics}
\label{sec:lpp}

In Sec.~\ref{sec:scalar-model}, we have derived corrections to the self-energy and coupling
constant at 1-loop level near to and far away from the event horizon. In this section, we study their physics.

\subsubsection{Particle physics: Near-horizon region}

We observe no physical particles in the locally Minkowski vacuum $|\Omega\rangle$
near the event-horizon region.
Indeed, one can speak about a \emph{massless} particle as a localised object in the near-horizon region 
when its de Broglie wavelength $\lambda_\mathbf{p}$ is much smaller than the size of the event horizon
$r_H$, i.e. $\lambda_\mathbf{p} \ll r_H$. More precisely, $\lambda_\mathbf{p}$ must actually be much
smaller than the size of a particle detector $l_D$ which is in turn much smaller than the horizon size. In
this case, the Wightman function $W_U(x,x')$ approximately coincides with the two-point
function as if there is no black hole
plus a small correction of the order of $(\lambda_\mathbf{p}/r_H)^2 \ll (\lambda_p/l_D)^2 \ll 1$.\footnote{To
our knowledge, this correction
which is due to the ``ingoing" part of the correlation function has not been discussed in the literature. In
the case of eternal black holes, this correction does not appear, because the decreasing (with the distance)
parts of the ``ingoing" and ``outgoing" modes cancel each other.} Therefore, the state  $|\Omega\rangle$ 
is not populated by the real particles. The same holds in the stationary frame near the
event horizon as the notion of particle is covariant.

There has been recently argued that an in-falling observer should discover a firewall (a sort of cloud of 
the high-energy (blue-shifted) Hawking particles $|\tilde{\psi}\rangle$) in the near-horizon 
region~\cite{Almheiri&Marolf&Polchinski&Sully}. If this is a real phenomenon, the equivalence principle 
does not hold, because the event horizon of evaporating black holes would then physically be a 
distinguishable set of space-time points. If so, the whole framework of general relativity which led to the 
notion of black hole would be not reliable. Importantly, no evidences have been found so far that the 
principle of equivalence does not hold.

The equivalence principle was sacrificed in favour of the unitarity~\cite{Almheiri&Marolf&Polchinski&Sully}. 
By the unitarity one should here understand the existence of the unitary $S$-matrix between the in-state
and the thermal out-state. The in-state in our notations is identified with the vacuum $|\Omega\rangle$,
while the out-state corresponds to the Boulware vacuum $|\tilde{\Omega}\rangle$. However, the mathematical
subtlety is that the in-state can only \emph{formally} be represented as a thermally populated state of the
Hawking particles defined with respect to the out-state. The relation between the in-state and out-state is
formal, because these do not define unitarily equivalent representations of the field operator
algebra~\cite{Emelyanov-2015-2}. Thus, this means that if one demands that there exists a unitary operator
$\hat{S}$ that relates the in-state and the thermally populated out-state, then one comes to the idea of
having the firewall near the event horizon. The problem is that the existence of the unitary operator $\hat{S}$
is not consistent with the principles of local quantum field theory as pointed out in~\cite{Emelyanov-2015-2}.

We have found above that the Feynman propagator $G_U(x,x')$ in the near-horizon region is given by a
thermal-like propagator with the imaginary ``temperature" $\Theta$ given in~\eqref{eq:imaginary-temperature}.
We interpret this small correction as being due to the modification of the field operator in the presence of
black holes. A similar effect occurs in the Casimir set-up. Indeed, if one considers a wave packet of a photon
of the de Broglie wavelength $\lambda_\mathbf{p} \ll d$, where $d$ is a distance between the conducting plates,
then the photon propagator turns out to be as a thermal-like one (for the modes in the perpendicular direction
with respect to the plates) with an imaginary ``temperature" $T_C = i/2d$ when localised far from the
plates~\cite{Latorre&Pascual&Tarrach}. The ``thermal" term in this propagator is understood as being due to
the boundary conditions satisfied by the electromagnetic operators or the vacuum fluctuations
which are present in the Minkowski vacuum.\footnote{The very existence of $T_C$ in the Casimir effect
can be envisaged from the Tomita-Takesaki theorem~\cite{Haag}. Indeed, according to the theorem 
the operator algebra composed of the electromagnetic field operators in-between the conducting plates must 
satisfy the Kubo-Martin-Schwinger condition in the Minkowski vacuum with respect to a certain one 
parameter group of automorphism of the quantum operators. This group can in general be of a geometrical 
as well as non-geometrical origin. In the present case, it is a symmetry related with the periodicity of the 
operators in the spatial direction which is transferred to the periodicity in the Minkowski time with a real 
period, which corresponds to the imaginary ``temperature". In a private discussion with Bernard Kay I got 
to know about a representation of the Minkowski vacuum as an ``imaginary-temperature state" in-between
the conducting plates, which was found in~\cite{Kay}.}

We want now to discuss the vacuum expectation value of the particle number operator in $|\Omega\rangle$.
This operator is defined as
\beqa\label{eq:number-operator-physical}
\hat{N}(h) &=& \hat{a}^\dagger(h)\hat{a}(h)\,,
\eeqa
where the wave packet $h(x)$ is the same as~\eqref{eq:wave-packet}, but for the scalar particle
and $h(p)$ having a maximum near the momentum $q$. Due to the thermal-like term of the imaginary
``temperature" $\Theta$ in $G_U(x,x')$, the (imaginary) quantity $\langle \hat{N}(h) \rangle$ does 
\emph{not} vanish. However, it is in general true that
the localised operators have a non-vanishing vacuum expectation value in the Minkowski vacuum
due to its Reeh-Schlieder property~\cite{Haag}.\footnote{As a consequence of this property, a 
sufficiently sensitive thermometer should measure a non-zero ``temperature" of the vacuum as being 
a local operator. For the same reason, the particle detector is excited all the time by ``particles"
in the vacuum. The ``temperature" and ``particles" of the vacuum are merely a quantum noise.} 
The physical interpretation of this mathematical theorem is given in terms 
of the quantum fluctuations. One needs thus to subtract these or calibrate the particle 
detector~\cite{Yngvason}. If we do the same in the near-horizon region, we obtain 
$\langle\hat{N}(h)\rangle = 0$.\footnote{We do not consider the imaginary
``temperature" $\Theta$ as being of any fundamental meaning,
rather than a footprint of our approach. We further study the near-horizon physics by employing
quantum kinetic theory in~\cite{Emelyanov-2017a}, wherein we derive the 2-point function
in the local inertial frame at $R \sim r_H$ without any reference to $\Theta$.}

\subsubsection{Particle physics: Far-horizon region}

The Feynman propagator $G_U(x,x')$ far away from the black hole ($R \gg r_H$) is on the contrary 
given by the propagator with a thermal-like term at the Hawking temperature $T_H$. This extra term
vanishes as $1/R^2$ at $R \rightarrow \infty$ and can be assigned to the modification of the field
operator in the presence of evaporating black holes.

Utilising the number operator introduced in Eq.~\eqref{eq:number-operator-physical}, one can define a 
number density operator. Its vacuum expectation value in $|\Omega\rangle$ for the ``outgoing" 
plane-wave modes within the frequency range from $\omega$ to $\omega +d\omega$ is given by
\beqa\
dn_\omega &=& \langle \hat{n}_\omega \rangle\,d\mu_\omega\,, \quad \text{where} \quad
\mu_\omega \;=\; \frac{\omega^2d\omega}{2\pi^2}
\eeqa
is the standard measure of integration, and the distribution of the modes in the frequency interval 
$(\omega,\omega + d\omega)$ is given by
\beqa\label{eq:number-local}
\langle \hat{n}_\omega \rangle &\approx& \frac{1}{4\omega^2R^2}\,\frac{\Gamma_\omega}{e^{\omega/T_H} - 1}\,,
\eeqa
where $\Gamma_\omega$ has been given in Eq.~\eqref{eq:gamma}.
It is worth pointing out that the right-hand side of~\eqref{eq:number-local} is fully due to
$\vec{W}_\beta(x,x')$ or the field operator $\hat{\Phi}_b(x) = \hat{b}(x) + \hat{b}^\dagger(x)$. 

The prefactor $1/4\omega^2R^2$ also appears in the \emph{effective} covariant Wigner function 
$\mathcal{W}_\text{eff}(x,p)$ playing a role of the phase-space distribution function in relativistic
kinetic theory (see, e.g.,~\cite{deGroot&vanLeeuwen&vanWeert}), namely
\beqa
\mathcal{W}_\text{eff}(x,k) &\approx& \frac{1}{(2\pi)^3}\frac{\Gamma_\omega}{4\omega^3R^2}\frac{\delta(\omega - k)}
{e^{\omega/T_H}-1} \;=\; f_\text{eff}(x,k)\,\frac{\delta(\omega - k)}{\omega}\,,
\eeqa
where $\omega = k_0 = k = |\mathbf{k}|$ with $\mathbf{k}^i = (k,0,0)$ and the index $i$ runs over
$\{r,\theta,\phi\}$. The effective function $f_\text{eff}(x,k)$ is known as the one-particle Wigner distribution. We
reproduce the result~\eqref{eq:number-local} by employing the  standard formula known in kinetic
theory:
\beqa
n &=& \int d^3p \,f_\text{eff}(x,p) \;=\; \int dn_\omega \;=\; \int d\mu_\omega \,\langle \hat{n}_\omega \rangle\,.
\eeqa
Moreover, we can compute the outward positive energy flux as found in~\cite{Christensen&Fulling,Candelas} 
as the second moment (with respect to the momentum) of the distribution
function~\cite{deGroot&vanLeeuwen&vanWeert}, namely
\beqa
\int d^3p\,p f_\text{eff}(x,p) &=& \frac{L}{4\pi R^2}\,, \quad 
\text{where} \quad L \;=\; \frac{1}{2\pi}\int d\omega\,\frac{\omega\,\Gamma_\omega}{e^{\omega/T_H} - 1}
\eeqa
is the luminosity. We further study local quantum physics near the event 
horizon by employing quantum kinetic theory in~\cite{Emelyanov-2017a}.

The equation~\eqref{eq:number-local}, however, differs from the distribution of the Hawking 
modes~\cite{Hawking}:
\beqa\label{eq:number-global}
\langle \hat{n}_\omega \rangle_H &=& \frac{\Gamma_\omega}{e^{\omega/T_H} - 1}\,.
\eeqa
By now the equation~\eqref{eq:number-global} is a widely-accepted result in black-hole physics.
This has been derived by computing the Bogolyubov coefficients relating $\hat{b}(x)$ with $\hat{a}_<(x)$ 
and $\hat{a}_<^\dagger(x)$~\cite{Hawking-1}. The reason of the discrepancy is that
the formula~\eqref{eq:number-local} is 
\emph{local} and expected to be valid only in a volume of the size being much smaller than $R \gg r_H$, 
whereas~\eqref{eq:number-global} is a global result. Thus, the physical meaning of~\eqref{eq:number-local}
and~\eqref{eq:number-global} is different. Specifically, the number of the locally plane-wave modes in a 
spherical shell of volume $dV = 4\pi R^2dR$ for the fixed distance $R \gg r_H$ from the black hole is given
by
\beqa
dN &=& n dV \;=\; 4\pi n R^2dR \;=\; 4\pi n R^2(dR/d\tau)d\tau \;=\; 4\pi n R^2d\tau\,,
\eeqa
where we have set $dR/d\tau = c = 1$. Therefore, we obtain the flux of these modes:
\beqa
\dot{N} &=& \frac{1}{2\pi}\int d\omega\,\langle \hat{n}_\omega \rangle_H\,
\eeqa
(or the number of the modes per the radial distance $dR$, i.e. $dN/dR$). It is worth noticing that the
number of these modes in a detector of volume $l_D{\times}l_D{\times}l_D$ drops out with the 
distance as $(l_D/R)^2$ for $R \gg r_H$. Analogously, we rederive a well-known result
\beqa
\dot{E} &=& \frac{1}{2\pi}\int d\omega\,\omega\,\langle \hat{n}_\omega \rangle_H
\eeqa
from the local distribution~\eqref{eq:number-local} by computing the energy in the spherical
shell of the volume $dV = 4\pi R^2dR$ or $dV = 4\pi R^2d\tau$. This implies that~\eqref{eq:correlator-far} 
and~\eqref{eq:propagator-far} are good approximations to the exact Wightman function and Feynman 
propagator whenever the conditions $|r-r'| \ll R$ and $r_H \ll R$ are satisfied.

All local observables are composed of the fundamental field operator $\hat{\Phi}(x)$. For instance, the
particle creation operator is given by Eq.~\eqref{eq:particle-creation-operator} and this is employed 
to construct the particle number operator. The field operator $\hat{\Phi}(x)$ can in turn be 
represented as the sum $\hat{\Phi}_>(x) + \hat{\Phi}_b(x)$ (see Eq.~\eqref{eq:scala-operator-splitting}). 
As emphasised above, the Wigner distribution $\mathcal{W}(x,k)$ is due to the part $\hat{\Phi}_b(x)$
of the field operator $\hat{\Phi}(x)$. In the asymptotically flat region, local physics is oblivious to the
presence of black holes, because the Wigner distribution drops out as $(r_H/R)^2$ for
$R \rightarrow \infty$. This means that if we choose the standard wave packet $h(x)$ in the spatial
infinity ($R \rightarrow \infty$) as we have been doing that on earth when we study scattering processes
in particle physics, we then find
\beqa
\hat{a}^\dagger(h) &=& \hat{a}_>^\dagger(h)\,,
\eeqa
where $\hat{a}_>^\dagger(x)$ is the standard creation operator of the scalar particle in Minkowski space-time
and the vacuum $|\Omega\rangle$ in the spatial infinity ``reduces" to the Minkowski 
vacuum (in the sense that the local operators probe $|\Omega\rangle$ at $R \rightarrow \infty$
as the Minkowski vacuum in Minkowski space-time), such that $\hat{a}_>(h)|\Omega\rangle = 0$. This 
implies that the \emph{effective} density matrix\footnote{Note that the description of the vacuum
$|\Omega\rangle$ in terms of the thermal density matrix comes from the assumption that the
probes of this vacuum in the spatial infinity ($R \rightarrow \infty$) are performed \emph{only} 
by the operators composed of $\hat{b}(x)$ and $\hat{b}^\dagger(x)$~\cite{Hawking}. However, this turns
out not to be the case, because $\hat{\Phi}(x)$ reduces to $\hat{\Phi}_>(x)$, rather than $\hat{\Phi}_b(x)$ in
the spatial infinity. The effective density matrix characterises the quantum
fluctuations of the $\hat{\Phi}_b$-part of the field $\hat{\Phi}(x)$, which is locally irrelevant at 
$R \gg r_H$.} introduced in black-hole physics must 
actually decrease with the distance as $(r_H/R)^2$, such that the vacuum $|\Omega\rangle$ is probed 
at $R \rightarrow \infty$ as being pure, but as if it is mixed at finite $R \gg r_H$. It implies that the formal 
representation of the vacuum $|\Omega\rangle$ as the thermally populated state of the particles defined 
with respect to $|\tilde{\Omega}\rangle$ in the spatial infinity is \emph{not} a self-consistent picture of 
the black-hole evaporation. 

Indeed, if we consider a wave packet $h(\mathbf{p}_n)$ (one of the elements of the countable set of 
orthonormalised functions) with a definite value of the momentum $\mathbf{p}_n$ 
localised in a ball of volume $\sigma \sim (\lambda_\mathbf{p})^3$ at the distance $R \gg r_H$ from 
an evaporating black hole, then the vacuum expectation value of the operator $\hat{N}(h)$
defined in Eq.~\eqref{eq:number-operator-physical} is given by
\beqa
\langle \hat{N}_{\mathbf{p}_n}\rangle &=& \frac{27r_H^2}{16R^2}\,\frac{1}{e^{\omega_n/T_H} - 1}
\;\rightarrow\; 0 \quad \text{for} \quad R \;\rightarrow\; \infty\,,
\eeqa
where $\omega_n = |\mathbf{p}_n|$. However, if we take a wave packet $h(\omega_n| lm)$ 
which corresponds to a spherical wave of the frequency $\omega_n$, 
the orbital number $l$ and the magnetic number $m$ localised around radial distance $R \gg r_H$, 
we obtain
\beqa
\langle \hat{N}_{\omega_n lm}\rangle &=& \frac{\Gamma_{\omega_n}}{e^{\omega_n/T_H} - 1}
\;\approx\; \frac{1}{4}\,\frac{27r_H^2\,\omega_n^2}{e^{\omega_n/T_H} - 1}\,.
\eeqa
This does \emph{not} depend on the radial distance $R$, because the support of the spherical shell
scales as $R^2$. Therefore, the Hawking particles are associated with the spherical waves. The
spherical waves are not localised in the angular directions. For this
reason, one cannot understand these as localised excitations, which one can put, for instance, in
a box of the one-cubic-meter size.

As shown above, the quantum operator $\hat{\Phi}(x)$ can be represented as $\hat{\Phi}_>(x) + \hat{\Phi}_b(x)$
above the horizon ($R > r_H$). The operators $\hat{\Phi}_>(x)$ and $\hat{\Phi}_b(x')$ commute with
each other for any points $x$ and $x'$, i.e.
\beqa
[\hat{\Phi}_>(x),\hat{\Phi}_b(x')] &=& 0\, \quad \text{for} \quad \forall\;\; x,x'\,.
\eeqa
In the asymptotically flat region, the operator $\hat{\Phi}_b(x')$ drops out to zero as $r_H/R$. It seems
this means that the matter outside of the hole is composed only of the operator $\hat{\Phi}_>(x)$
at $R \rightarrow \infty$ and \emph{cannot} be used to \emph{directly} discover the Hawking modes
which are due to $\hat{\Phi}_b(x)$. It is still possible to discover these indirectly through its gravitational 
influence, because $\langle \hat{T}_{\nu}^{\mu}\rangle \neq 0$. Thus, these could be a sort of ``the dark 
radiation". Another argument in favour of this idea is the following: If we prepare a thermal gas in a small 
box at $R \rightarrow \infty$ and let it fall towards the horizon, then the energy-momentum tensor will be 
finite at $R = r_H$ within the box. This is \emph{not} the case for the Hawking gas, because the 
energy-momentum tensor will diverge for any temperature $T \neq T_H$ on the horizon and the whole
consideration becomes self-inconsistent (unless one starts to treat this gedankenexperiment at the level 
of non-perturbative quantum gravity). Therefore, one might conclude that ``the Hawking
matter" is decoupled from 
the normal matter. However, this statement makes sense only if one assumes that the 
``outgoing" modes correspond to real particles which can be literally used to prepare the Hawking 
gas heated up to any temperature $T$. This decoupling does not make any sense, if one understands 
these as virtual particles/vacuum fluctuations, because the fundamental field is $\hat{\Phi}(x)$, rather than 
$\hat{\Phi}_>(x)$ or $\hat{\Phi}_b(x)$ separately.

If we calibrate the particle detector (as made above at $R \sim r_H$) or subtract the vacuum contribution to 
the Wigner function at large, but fixed $R$, then we obtain $|\Omega\rangle$ is empty. It does 
not imply the vanishing energy-momentum tensor, i.e. $\langle \hat{T}_\nu^\mu \rangle \neq 0$.
For the same reason, the Fourier transform of the Wightman function with respect to the time (this 
represents the Unruh-DeWitt detector) is also non-vanishing (although $|\Omega\rangle$ does not 
contain the physical particles), because this yields the frequency spectrum of the vacuum fluctuations 
(according to the Wiener-Khinchin theorem) as pointed out in~\cite{Candelas}.

Since we have been trying to interpret the black-hole evaporation as the vacuum would possess
(inhomogeneous, but isotropic) medium-like properties (because this interpretation seems to be 
self-consistent and does not suffer from the absence of the unitary $S$-matrix as understood by 
many researchers as well as the firewall problem), it is of interest to be aware of any other physical 
examples when this kind of the viewpoint is fruitful. There exists at least one to our 
knowledge. Specifically, the propagation of photons in the Minkowski vacuum with a super-strong 
magnetic field $B$ ($\,\gtrsim \pi m_e^2/\alpha e$, where $\alpha$ is the fine structure constant 
and $e$ the elementary charge) occurs as if the photons move through a magnetised physical 
plasma, i.e. in the plasma held at the external magnetic field~\cite{Melrose&Stoneham,Dittrich&Gies}.

\section{Concluding remarks}
\label{sec:concluding remarks}

\subsection{Particles in black-hole geometry}

We have proposed a new, covariant definition of the notion of particle in curved space-time
which is observer-independent. This definition is mostly motivated by the success of particle physics we have
been testing on earth and is consistent with the particle creation effect in expanding universe~\cite{Parker}
(see for a recent short review~\cite{Parker-1}).

We have found that the term in the 2-point function that is (partially) responsible for the black-hole
evaporation is of the sub-leading
order with respect to the term providing the correct pole structure of the Feynman propagator. Namely,
this hierarchy of the terms is regulated by the ratio $(\lambda_\mathbf{p}/r_H)^2$ near the horizon, where
$\lambda_\mathbf{p}$ is a de Broglie wavelength of the scalar particle of momentum $\mathbf{p}$. The
modes leading to this correction should thus be understood as vacuum fluctuations. In the far-horizon region,
this suppression is even stronger: $(\lambda_\mathbf{p}/R)^2$ for $R \gg r_H$. This implies that the black-hole
evaporation originating in the near-horizon region cannot be understood as a local effect that agrees
with~\cite{Bardeen}.

The flux of the energy density changes its direction at the distance $R \sim 3M$ outside of the
event horizon~\cite{Unruh-1,Giddings}. Hence, it might imply that the sub-leading term in the propagator
is associated with the flux of the negative energy density inside the black-hole horizon
(see~\cite{Emelyanov-2017a}). Thus, we come to a conclusion that the Hawking's partner
mode is a vacuum fluctuation. The same observation based on a different argument has been
recently made in~\cite{Hotta&Schuetzhold&Unruh}.

\subsection{One-loop correction to self-energy}

The tadpole diagram yields the effective mass of the self-interacting scalar field $\Phi(x)$. Employing
the effective action approach for computing the 1-loop correction to the scalar field equation, one can
obtain the well-known result $m_\Phi^2 = \frac{\lambda}{2}\langle\hat{\Phi}^2(x)\rangle$. The Wick squared
operator $\hat{\Phi}^2(x)$ in the Unruh vacuum was derived in~\cite{Candelas}.
Hence, the value of the effective scalar mass at one-loop approximation is a straightforwardly computable
quantity.

Perhaps, the less trivial computation is to reproduce the result for $m_\Phi^2$ by applying Feynman's 
method in the freely-falling frame. The non-trivial part of this computation is how to ``properly" renormalise
the ultraviolet divergence of the tadpole diagram. As a guideline, we have used the above mentioned
findings for the Wick squared operator. 
This can be considered as an intermediate step to compute other scattering reactions
with the radiative corrections.

\subsection{One-loop correction to coupling constant}

A presence of black holes entails the modification of quantum field propagators. This
in turn leads to the non-trivial corrections to the self-energy and coupling constant in the loop
expansion. The latter gets a loop contribution depending on the external momenta $q \gg \kappa$. 

We have found that the 1-loop correction to $\lambda(q)$ in the near-horizon region 
$R \sim r_H$ is larger than that in the absence of the black hole, while this reduces to the standard 
result found in the Minkowski-space approximation in the asymptotically flat region
$R \gg r_H$.\footnote{We have derived in~\cite{Emelyanov-2017a} the higher-order corrections
to the 2-point function in $\Delta\mathbf{x}$ up to the second order in both the far-horizon and
near-horizon region. These corrections give a contribution to the running coupling constant
$\lambda(q)$ of the order of $(1+3\cos^2\gamma)\,T_H^4/q^4$, where $\gamma$ is an angle between
$\mathbf{q}$ and $\mathbf{n} = \mathbf{R}/R$. This is much smaller than the
leading order term found above in the regime $q \gg T_H$.}

\subsection{One-loop correction to vacuum energy}

The energy-momentum tensor of the scalar field gets a correction due to the self-coupling. 
The one-loop contribution to the vacuum energy is given by the standard method of the
perturbation theory, namely
\beqa\label{eq:loc-emt}
\langle\Delta\hat{T}_{\mu\nu}\rangle &=& \frac{\lambda}{4!}\,\eta_{\mu\nu}\langle\hat{\Phi}^4\rangle \;=\;
\frac{\lambda}{8}\,\eta_{\mu\nu}\langle\hat{\Phi}^2\rangle^2\,.
\eeqa
The same result can be reproduced through taking into account vacuum bubbles at two-loop level,
namely
\beqa\nonumber\label{eq:emt-one-loop-correction}
\langle\Delta\hat{T}_{\mu\nu}\rangle &=& \frac{i}{8V_4}\,\eta_{\mu\nu}\left(
\mathbf{\imineq{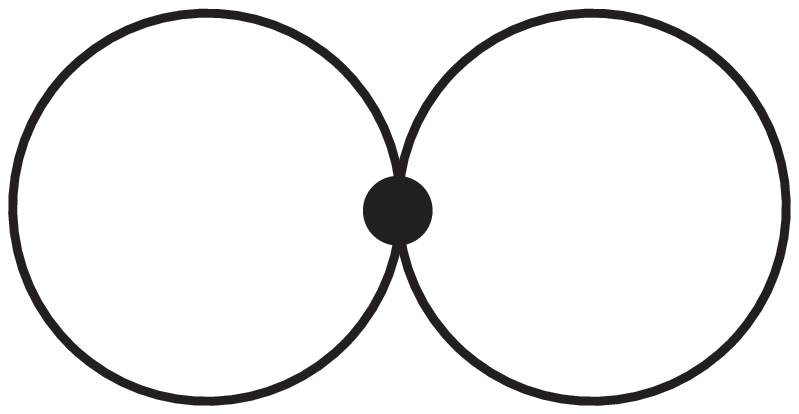}{6.0}} + 
2\mathbf{\imineq{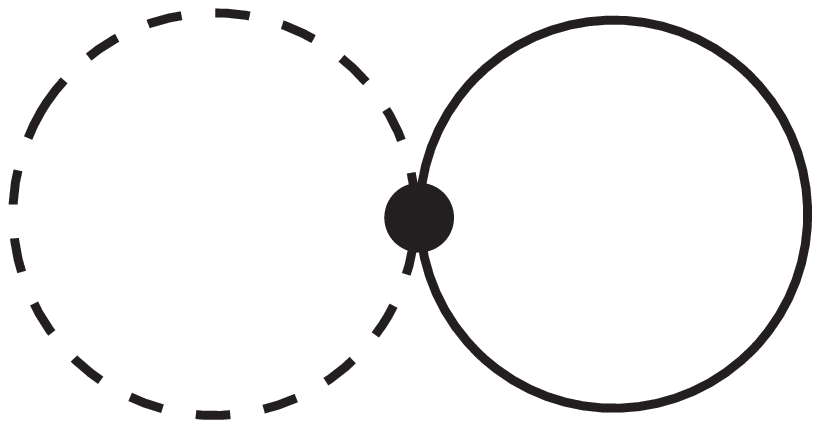}{6.0}} +
\mathbf{\imineq{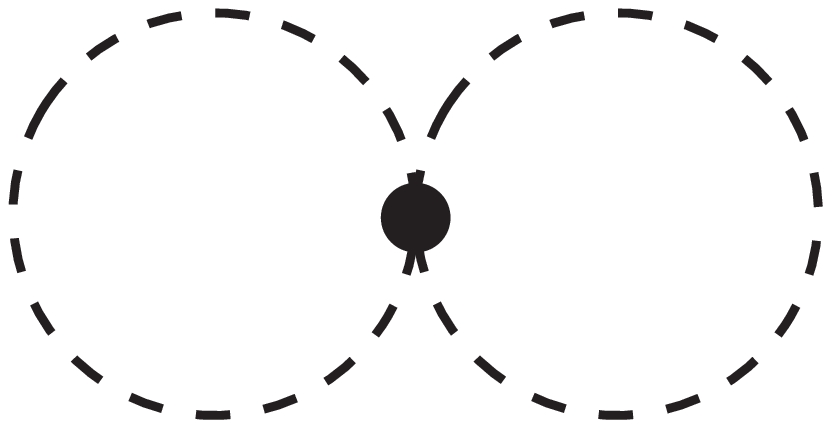}{6.0}}\right)
\\[2mm]
&=& \frac{\lambda}{8}\,\eta_{\mu\nu}\Big(G_U(x,x) - G_H(x,x)\Big)^2 \;=\; 
\frac{\lambda}{8}\,\eta_{\mu\nu}\langle\hat{\Phi}^2\rangle^2\
\eeqa
according to Eqs. \eqref{eq:smws} and \eqref{eq:effective-mass-diagram},
where $V_4 = \int d^4x$ is a four-dimensional volume of
a local Minkowski frame as if it is infinitely large.

Thus, we find that the 2-loop correction to the vacuum energy after having been renormalised is finite 
at $R \sim r_H$ in the freely-falling frame. In the asymptotically flat region, the 2-loop correction to the
vacuum energy is of the order of $1/R^4$. This is much smaller than the 1-loop contribution that
vanishes as $1/r_H^2R^2$ in the limit $R \rightarrow \infty$~\cite{Christensen&Fulling,Candelas}.

\subsection{Local renormalisation scheme}

A sort of ambiguity is inherent to the local renormalisation scheme we have proposed. One could choose 
the fictitious field $\phi(x)$ be anticommuting (instead of commuting) and with a \emph{positive} norm
(with the correct sign in front of the propagator). This would change only the 1-loop correction to the 
coupling constant which depends on the external momenta, while \eqref{eq:effective-mass-lrs} and 
\eqref{eq:emt-one-loop-correction} are insensitive to these modifications. Our choice is, however, 
symmetric with respect to how the propagators $G_U(x,x')$ and $G_H(x,x')$ contribute to the coupling 
constant $\lambda$ (at least) at one-loop level.

In addition, the coupling constant $\lambda$ in the thermal state of temperature $T$ without introducing 
the fictitious field $\phi(x)$ would be
\beqa
\lambda(q,T) &=& \lambda + \frac{3\lambda^2}{32\pi^2}\left(\ln\Big(\frac{q^2}{4\pi \mu^2}\Big) 
- \frac{2\pi^2}{3q^2}\,T^2\right) + \text{O}\big(\lambda^3,\lambda^2T^4/q^4\big).
\eeqa
Thus, the temperature-dependent correction is double of what one finds if the fictitious field is taken 
into account. However, it will be the same result if we also heat the fictitious field up to the temperature $T$.

In the case of a massive $\lambda\Phi^4$-theory, the one-loop correction to the scalar self-energy
after having been regularised by one of the standard methods (dimensional or Pauli-Villars regularisation) 
is non-vanishing after the subtraction of the UV-divergent term. However, the extra UV-divergent term
($\,\propto m^2\ln\sigma_0(x,x')$) of $\langle\hat{\Phi}(x)\hat{\Phi}(x')\rangle$ in the limit $x' \rightarrow x$ is
precisely cancelled by the same term in the Hadamard parametrix. Thus, the mass of the scalar field
does not get a quantum correction at one-loop level in this renormalisation approach.

The UV-divergent part of the stress tensor in Minkowski space is associated with the
vacuum bubbles. One usually ignores this in particle physics as these do not show up for scattering
processes (where one measures the energy differences only). This is illegitimate when one takes 
gravity into account. One of the methods to compute the renormalised stress tensor is to subtract 
the Hadamard parametrix from the Wightman function (see, e.g.,~\cite{Decanini&Folacci} for a brief 
review). 

It is a well-known problem in standard electroweak theory, that the quantum/loop contributions to
the self-energy of the Higgs field are (polynomially) divergent. That is the origin of the hierarchy
problem. In this case, the one-loop self-energy term depends on the external momentum of the
Higgs particle. If we do not introduce the fictitious field in the loop diagrams containing the external
momenta, then we recover the standard results.

To summarise, we should either introduce the fictitious field only for the purpose to renormalise 
expectation values of local quantum operators in Hadamard states or demand that the fictitious field 
also appears in the loop diagrams depending on the external momenta. The consequences of the 
latter should however be investigated in detail to draw any decisive conclusions.

\subsection{Locally Minkowski and Unruh vacuum}

These vacua are indistinguishable in the far-from-horizon region at the leading order of the
approximation. The deviation of the Unruh vacuum from the locally Minkowski one $|\Omega\rangle$
can be, however, established near the event horizon, taking into account that these vacua are
characterised by the 2-point functions $W_U(x,x')$ and $W(x,x')$, respectively.

The singular part of $W_U(x,x')$ at $r \sim r_H$ is given by $\vec{W}_{\beta}(x,x')$ (see
Eq.~\eqref{eq:unruh-2-point-function}). It is inversely proportional to
$\cosh(\kappa\Delta{t}_S) - \cosh(\kappa(2\bar{\sigma}(\mathbf{x},\mathbf{x}'))^\frac{1}{2})$, where
$\kappa = \frac{1}{2}f'(r_H)$. This can be transformed to the Minkowski form $\sigma_0(x,x')$ following
the procedure outlined in Sec.~\ref{sec:olctse-nhr}, where, e.g., the new time coordinate is approximately
equal to $f^\frac{1}{2}(r)\sinh(\kappa t_S)/\kappa$.

The singular part of $W(x,x')$ is approximately given by $-1/(8\pi^2\sigma(x,x'))$, where $\sigma(x,x')$
is a geodetic distance between the space-time points $x$ and $x'$. This can be expressed in terms of
$\sigma_0(x,x') \propto \cosh(\kappa\Delta{t}_S) - \cosh(\kappa(2\bar{\sigma}(\mathbf{x},\mathbf{x}'))^\frac{1}{2})$
as found in Eq.~\eqref{eq:gdnh}. However, in terms of the Riemann normal coordinates,
$\sigma(x,x')$ acquires a Minkowski form as well, but at a fixed spatial point at $r \sim r_H$, the Riemann
normal time reads
\beqa
\Delta{y}^0 &\approx& \frac{f^\frac{1}{2}(r)}{\frac{1}{2}f'(r)}\,\sinh\left(\frac{1}{2}f'(r)\Delta{t}_S\right).
\eeqa
Thus, the Unruh vacuum differs from the locally Minkowski vacuum in the near-horizon region,
because $\kappa = \frac{1}{2}f'(r_H)$ and $\frac{1}{2}f'(r)$ coincide in the limit $r \rightarrow r_H$
(implying $f(r) \rightarrow 0$) only. We shall determine $W(x,x')$ at any radial distance from the event horizon
elsewhere.

\section*{
ACKNOWLEDGMENTS}
I am thankful to Frans Klinkhamer, Jos\'{e} Queiruga and Frasher Loshaj for discussions and their 
comments on an early version of this paper. It is also a pleasure to thank Eduardo Grossi for the
reference~\cite{deGroot&vanLeeuwen&vanWeert}.


\begin{thebibliography}{99}

\bibitem{Hawking}
S.W. Hawking,
\hspace*{0mm}``Breakdown of predictability in gravitational collapse,''
Phys. Rev. D{\bf 14}, 2460 (1976).

\bibitem{Almheiri&Marolf&Polchinski&Sully}
A. Almheiri, D. Marolf, J. Polchinski, J. Sully,
\hspace*{0mm}``Black holes: complementarity or firewalls?,''
JHEP02, 062 (2013), arXiv:hep-th/1207.3123.

\bibitem{Emelyanov-2015-2}
S. Emelyanov,
\hspace*{0mm}``Can gravitational collapse and black-hole evaporation be a unitary process after all?,''
arXiv:hep-th/1507.03025.

\bibitem{Christensen&Fulling}
S.M. Christensen, S.A. Fulling
\hspace*{0mm}``Trace anomalies and the Hawking effect,''
Phys. Rev. D{\bf 15}, 2088 (1977).

\bibitem{Candelas}
P. Candelas,
\hspace*{0mm}``Vacuum polarization in Schwarzschild spacetime,''
Phys. Rev. D{\bf 21}, 2185 (1980).

\bibitem{Scharnhorst}
K. Scharnhorst,
\hspace*{0mm}``On propagation of light in the vacuum between plates,''
Phys. Lett. B{\bf 236}, 354 (1990).

\bibitem{Emelyanov-2016-1}
S. Emelyanov,
\hspace*{0mm}``Low-energy electromagnetic radiation as an indirect probe of black-hole evaporation,''
Nucl. Phys. B{\bf 913}, 318 (2016), arXiv:hep-th/1602.01475.

\bibitem{Emelyanov-2016-2}
S. Emelyanov,
\hspace*{0mm}``Effective photon mass from black-hole formation,''
Nucl. Phys. B{\bf 919}, 110 (2017), arXiv:hep-th/1603.01148.

\bibitem{Mukhanov}
V.F. Mukhanov,
\hspace*{0mm}{\sl Physical Foundations of Cosmology}
(Cambridge University Press, 2005).

\bibitem{Haag}
R. Haag,
\hspace*{0mm}{\sl Local quantum physics. Fields, Particles, Algebras}
(Springer-Verlag, 1996).

\bibitem{Martel&Poisson}
K. Martel, E. Poisson,
\hspace*{0mm}``Regular coordinate systems for Schwarzschild and other 
spherical spacetimes,''
Am. J. Phys. {\bf 69}, 476 (2001), arXiv:gr-qc/0001069.

\bibitem{Unruh}
W.G. Unruh,
\hspace*{0mm}``Notes on black hole evaporation,''
Phys. Rev. D{\bf 14}, 870 (1976).

\bibitem{DeWitt}
B. DeWitt,
\hspace*{0mm}``Quantum field theory in curved spacetime,''
Phys. Rep. {\bf 19}, 295 (1975).

\bibitem{DeWitt}
B. DeWitt,
\hspace*{0mm}{\sl The global approach to quantum field theorie}
(V2, Oxford University Press, 2003).

\bibitem{Boulware}
D.G. Boulware,
\hspace*{0mm}``Quantum field theory in Schwarzschild and Rindler spaces,''
Phys. Rev. D{\bf 11}, 1404 (1975).

\bibitem{Emelyanov-2014-2}
S. Emelyanov,
\hspace*{0mm}``Non-unitarity or hidden observables?,''
arXiv:gr-qc/1410.6149.

\bibitem{Emelyanov-2015-1}
S. Emelyanov,
\hspace*{0mm}``Observing quantum gravity in asymptotically AdS space,''
Phys. Rev. D{\bf 92}, 124062 (2015), arXiv:hep-th/1504.05164.

\bibitem{Moretti}
V. Moretti,
\hspace*{0mm}``Comments on the stress energy tensor operator in curved space-time,''
Commun. Math. Phys. {\bf 232}, 189 (2003), arXiv:gr-qc/0109048.

\bibitem{Decanini&Folacci}
Y. D\'ecanini, A. Folacci,
\hspace*{0mm}``Hadamard renormalization of the stress-energy tensor for a quantized scalar field 
in a general spacetime of arbitrary dimension,'' Phys. Rev. D{\bf 78}, 044025 (2008), arXiv:gr-qc/0512118.

\bibitem{Kleinert&Frohlinde}
H. Kleinert, V. Schulte-Frohlinde,
\hspace*{0mm}{\sl Critical properties of $\phi^4$-theories},
(World Scientific Publishing, 2001).

\bibitem{Latorre&Pascual&Tarrach}
J.I. Latorre, P. Pascual, R. Tarrach,
\hspace*{0mm}``Speed of light in non-trivial vacua,''
Nucl. Phys. B{\bf 437}, 60 (1994), arXiv:hep-th/9408016.

\bibitem{Kay}
B.S. Kay,
\hspace*{0mm}``Casimir effect in quantum field theory,''
Rev. Phys. D{\bf 20}, 3052 (1979).

\bibitem{Yngvason}
J. Yngvason,
\hspace*{0mm}``The role of type III factors in quantum field theory,''
Rep. Math. Phys. {\bf 55}, 135 (2004), arXiv:math-ph/0411058.

\bibitem{Emelyanov-2017a}
S. Emelyanov,
\hspace*{0mm}``Quantum kinetic theory of a massless scalar model in the presence of a
Schwarzschild black hole,'' Annalen der Physik, 1700078 (2017), arXiv:hep-th/1703.01674.

\bibitem{deGroot&vanLeeuwen&vanWeert}
S.R. de Groot, W.A. van Leeuwen, C.G. van Weert,
\hspace*{0mm}{\sl Relativistic Kinetic Theory},
(North-Holland, 1980).

\bibitem{Hawking-1}
S.W. Hawking,
\hspace*{0mm}``Black hole explosions?,'' Nature {\bf 248}, 30 (1974);
``Particle creation by black holes,'' Commun. Math. Phys. {\bf 43}, 199 (1975).

\bibitem{Melrose&Stoneham}
D.B. Melrose, R.J. Stoneham,
\hspace*{0mm}``Vacuum polarization and photon propagation in a magnetic field,''
Nuovo Cim. A{\bf 32}, 435 (1976).

\bibitem{Dittrich&Gies}
W. Dittrich, H. Gies,
\hspace*{0mm}{\sl Probing the quantum vacuum. Perturbative effective action
approach in quantum electrodynamics and its application},
(Springer-Verlag, 2000).

\bibitem{Parker}
L. Parker,
\hspace*{0mm}``Particle creation in expanding universe,''
Phys. Rev. Let. {\bf 21}, 562 (1968);
\hspace*{0mm}``Quantized fields and particle creation in expanding universe. I,''
Phys. Rev. {\bf 183}, 1057 (1969);
\hspace*{0mm}``Quantized fields and particle creation in expanding universe. II,''
Phys. Rev. D{\bf 3}, 346 (1971).

\bibitem{Parker-1}
L. Parker,
\hspace*{0mm}``Creation of quantized particles, gravitons and scalar perturbations by the expanding universe,''
J. Phys. Conf. Ser.  {\bf 600}, 012001 (2015); arXiv:gr-qc/1503.00359.

\bibitem{Bardeen}
J. Bardeen,
\hspace*{0mm}``Black hole evaporation without an event horizon,''
arXiv:gr-qc/1406.4098.

\bibitem{Unruh-1}
W.G. Unruh,
\hspace*{0mm}``Origin of the particles in black-hole evaporation,''
Phys. Rev. D{\bf 15}, 365 (1977).

\bibitem{Giddings}
S.B. Giddings,
\hspace*{0mm}``Hawking radiation, the Stefan-Boltzmann law, and unitarization,''
Phys. Lett. B{\bf 754}, 39 (2016); arXiv:hep-th/1511.08221.

\bibitem{Hotta&Schuetzhold&Unruh}
M. Hotta, R. Sch\"{u}tzhold, W.G. Unruh,
\hspace*{0mm}``Partner particles for moving mirror radiation and black hole evaporation,''
Phys. Rev. D{\bf 91}, 124060 (2015), arXiv:gr-qc/1503.06109.

\end{thebibliography}
\end{document}